\documentclass[final,3p,times,twocolumn]{elsarticle}

\usepackage{lineno}

\usepackage{amssymb}
\usepackage{amsmath}
\usepackage{bm}
\usepackage{float}
\usepackage{subcaption}
\usepackage{amsthm}
\usepackage{flushend}
\usepackage{url}

\journal{Results in Physics}

\begin{document}

\begin{frontmatter}



\title{Superradiant pulse saturation in a Free Electron Laser}


\author[1,2,3]{Pornthep Pongchalee\corref{cor1}}
\ead{pornthep.pongchalee@strath.ac.uk}

\author[1,2]{Brian W.J. M$^{\rm c}$Neil}

\affiliation[1]{organization={Department of Physics, University of Strathclyde},
            city={Glasgow},
            postcode={G4 0NG}, 
            country={UK}}
\affiliation[2]{organization={Cockcroft Institute},
            city={Warrington},
            postcode={WA4 4AD}, 
            country={UK}}
            
\affiliation[3]{organization={ASTeC, STFC Daresbury Laboratory},
            city={Warrington},
            postcode={WA4 4AD}, 
            country={UK}}
\cortext[cor1]{Corresponding author}
            
\begin{abstract}
A study is made of the saturation mechanism of a single superradiant `spike' of radiation in a Free Electron Laser. A one-dimensional (1D) computer model is developed using the Puffin, un-averaged FEL simulation code, which allows sub-radiation wavelength evolution of both the spike radiation field and the electron dynamics to be modelled until the highly non-linear saturation process of the spike is observed. Animations of the process from the start to the end of the interaction are available. The resultant saturated spike duration is at the sub-wavelength scale and has a broad spectrum. The electrons passing through the spike can both lose and gain energy many times greater than that of the `normal' non-pulsed FEL interaction. A saturation mechanism is proposed and tested via a simple analysis of the 1D FEL equations. The scaling results of the analysis are seen to be in good agreement with the numerical results. A simple model of three dimensional diffraction effects of the radiation is applied to the results of the 1D simulations. This greatly reduces longer wavelengths of the power spectrum, which are seen to be emitted mainly after the electrons have propagated through the spike, and is seen to be in qualitative agreement with recent experimental results.

\end{abstract}




\end{frontmatter}


\section{Introduction}
\label{intro}
Analytical studies of FEL amplifiers have identified two distinct solutions to the equations describing the co-propagation of the electrons and radiation in the FEL undulator: the Steady-State and Superradiant regimes~\cite{ BONIFACIO198536,mcneil1988NIMA, PhysRevA.40.4467, bonifacio1988slippage, atoms9020028}. When the equations describing the  FEL electron--radiation  interaction are averaged over at least one resonant radiation wavelength, the pulsed superradiant emission, describing pulse effects, has been shown to have a hyperbolic secant solution for the radiation field emitted in this regime~\cite{piovella1991hyperbolic}. An experiment has also recently investigated superradiant pulses obtained from an FEL oscillator~\cite{zen2023full}.

As a superradiant radiation  pulse, or `spike', evolves in an electron beam, the averaged analysis and numerical simulations predicts that as the spike's peak power increases,  its temporal duration decreases. Clearly, both the analysis and  simulations will start to break down when spike durations start to approach that of the radiation wavelength.  Hence, the evolution of ultra-short spikes in FELs is not fully described or understood. Both analytical methods and numerical simulations are unable to determine whether superadiant spikes eventually reach a saturation point and the self-similar solution breaks down. A preliminary study has already shown that the averaged and unaveraged numerical simulations diverge, once sub-period effects occur in the evolution of ultra-short superadiance FEL spikes~\cite{mcneil2003improved, campbellnjp}.

To help advance our understanding of the spike evolution, the unaveraged numerical simulation code Puffin~\cite{campbell2012puffin, campbell2018updated} was used to study the highly nonlinear FEL radiation spike growth and evolution as it propagates through a uniform, effectively infinitely long, electron beam. These simulations were conducted in the 1D, cold beam limit, and have unveiled a novel regime in which the superradiant spike is seen to saturate. This research provides an idealised baseline and gives new insights into the evolution of FEL superradiant spikes and the electron dynamics at sub-wavelength scales.

\section{Simulation model}
Unaveraged numerical simulations of FEL pulse evolution using Puffin in the 1D limit are now presented. The parameters used in the simulation are not intended to represent any particular existing or proposed FEL experiment, but are used to investigate the basic FEL interaction as it enters the highly non-linear regime of very intense radiation pulse evolution with sub-wavelength resolution. Previous studies comparing Puffin with experiments and averaged 3D FEL simulation codes have shown good agreement when the FEL parameters do not vary significantly over a radiation wavelength~\cite{campbellnjp}. 

The parameters used here are scaled using the FEL parameter $\rho$, defined as~\cite{BONIFACIO1984373}:
\begin{equation}
    \rho = \frac{1}{\gamma_r}\left( \frac{a_w \omega_p}{4 c k_w} \right)^{2/3},
\end{equation}
where $\gamma_r$ is the resonant electron beam Lorentz factor, $a_w$ is the undulator parameter, $k_w = 2\pi/\lambda_w$ is the undulator wavenumber, $\omega_p = \sqrt{e^2 n_p/\epsilon_0 m_e}$ is the non-relativistic electron beam plasma frequency, and $n_p$ is the peak number density of the electron beam.  

Other important scaling parameters derived from this fundamental scaling parameter include~\cite{mcneil2010x}: the cooperation length, $l_c = \lambda_r/4\pi\rho$, where $\lambda_r$ is the resonant radiation wavelength; the gain length, $l_g =  \lambda_w/4 \pi\rho$, where $\lambda_w$ is the undulator period; the scaled distance through the undulator $\bar{z} = z/l_g= 4\pi\rho N_w$, where $N_w$ is the number of the undulator periods; and $\bar{z}_2 = (ct-z)/l_c$ is the scaled length in the radiation frame of reference.

\begin{figure*}
\centering
\includegraphics[width=0.90\textwidth]{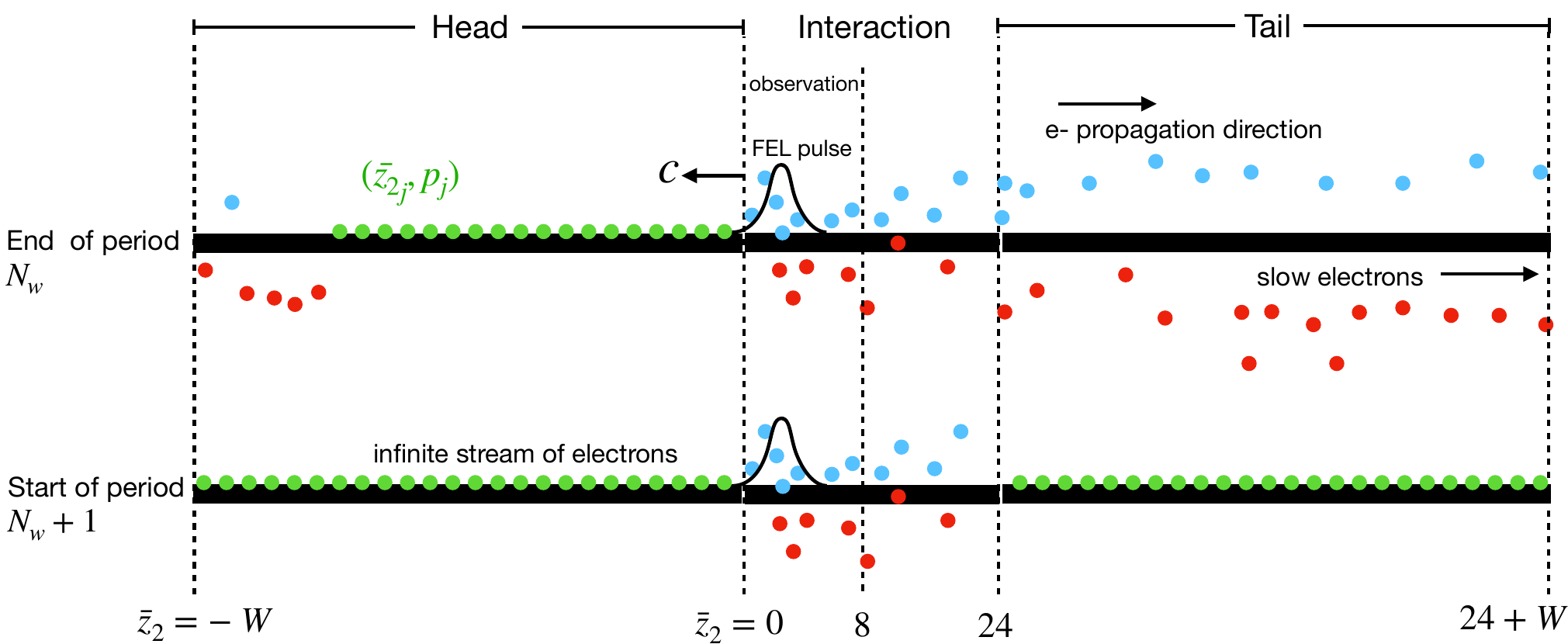}
    \caption{Schematic of the simulation window in the scaled radiation frame of reference $\bar{z}_2 = (ct-z)/l_c$. It consists of three main regions, Head, Interaction and Tail. All macroparticles propagate left-to-right as their speed is less than that of light. Top: The macroparticles representing the electron beam have propagated one undulator period (left-to-right) through an intense radiation pulse. The `slow' macroparticles (red) have lost energy to the radiation pulse and some have  propagated into the tail window $24<\bar{z}_2<24+W$, as have some that have gained energy from the pulse (blue). Those that propagate further than the Tail window, $\bar{z}_2>24+W$, are re-assigned into the Head window by application of periodic boundary conditions over $-W<\bar{z}_2<24+W$. Bottom: All macroparticles in both the Head and Tail windows are then re-initialised as shown, with equal spacing in $\bar{z}_2$ and a resonant, monoenergetic distribution. Here $W$ represents the range of the windows in $\bar{z}_2$ for simulating different values of $\rho$, e.g. $W=238$ for $1 \leq 4\pi\rho \leq 1.25$, and $W=488$ for $0.5 \leq 4\pi\rho < 1$.}
\label{fig:windows}
\end{figure*}

In the Puffin simulation code, the electron beam is modelled using a system of macroparticles. The beam is assumed to be an infinitely long `CW' cold beam, i.e. zero energy spread, with all initial parameters having a uniform distribution. The coupled radiation/electron equations are solved in a simulation window travelling at the speed of light and therefore of fixed width in $\bar{z}_2$. As the electrons propagate at a lesser speed, the macroparticles therefore always propagate to larger values of $\bar{z}_2$ in this window and with a resonant macroparticle having a propagation speed of $d\bar{z}_2 / d\bar{z}=1$. The fact that $d\bar{z}_2 / d\bar{z}>0$ for all macroparticles assists in the modelling of an infinite electron beam. The simulation window is defined by three separate regions: the head, interaction and tail windows. The interaction window contains the initial radiation pulse field and is where the spike of the radiation evolves. The macroparticle positions in $\bar{z}_2$ are tested at the end of each undulator period and those that have propagated outside of the interaction window, into the tail window, are re-initialised in both position and energy to re-populate the head window and so maintain a CW electron beam. The initial number of macroparticles per radiation wavelength is set to 800.
A schematic of the simulation windows and the process of re-initialisation of the macroparticles between at the end of $N_w$ and before starting $N_w+1$ undulator periods is shown in figure~\ref{fig:windows}.

The initial electron phases of the $j$th macroparticle, $\bar{z}_{2_j}$, are uniformly distributed over the simulation window, with all Lorentz  factors set equal to the resonant energy $\gamma_r=100$, so that $p_j = (\gamma_j-\gamma_r)/\rho\gamma_r = 0$ $\forall j$. The undulator has period $\lambda_w = 4~\text{cm}$ and undulator parameter $a_w = 1.0$.

\section{Simulation Example}
Superradiant pulse behaviour into the highly non-linear regime and with sub-wavelength resolution is now investigated using the above simulation method and parameters in a helical undulator configuration. 
The radiation-electron evolution is first modelled from a relatively low power input seed pulse through to a very short, high power superradiant spike.  It is demonstrated that, for the relatively large value of $\rho=1/4\pi$ used, the superradiant spike interaction saturates following a long propagation distance of approximately 400 gain lengths $(\bar{z}\approx 400)$ and with a peak scaled intensity approximately 4000 times the usual steady-state saturation intensity of $|A_{sat}|^2 \sim 1.4$~\cite{BONIFACIO1984373}. While this type of evolution may not currently be practical with current FEL systems, it is of general interest to observe the radiation spike saturation process.

The scaling of the saturated values of peak spike energy etc, is then investigated for different values of $\rho$, and an estimate is made of how 3D diffractive effects may affect the radiation spike properties.

\subsection{Superradiant pulse simulation}
With the value of $\rho=1/4\pi$, one radiation period corresponds to a scaled length of $\Delta\bar{z}_2 = 1$, and one undulator period of $\Delta{\bar{z}} = 1$.
A relatively small Gaussian seed radiation  field of scaled intensity $|A_0|^2 = 0.4$, width $\sigma(\bar{z}_2) = 2$ and centred at $\bar{z}_2 = 12$, is injected into the interaction window  $0 \leq \bar{z}_2 \leq 24$ and is set to zero elsewhere at the undulator entrance $\bar{z} = 0$. The field is sampled uniformly at $201$ points per radiation wavelength.  After propagating a scaled distance of $\bar{z} = 4 \pi\rho N$ for $N = 1,2,3,...,N_w$, where $N_w$ is the number of undulator periods, the macroparticles and the radiation field outside the interval $0 \leq \bar{z}_2 \leq 24$ are re-initialised in the head and tail windows as described above. The number of integration steps over one undulator period is $800$, with the macroparticle  and radiation data saved every $20$ steps. The sub-wavelength radiation and electron dynamics can then be observed with the resolution of $\lambda_r/20$. 

\begin{figure*}
    \centering
    \begin{subfigure}[c]{0.49\textwidth}
        \includegraphics[trim=1.5cm 1cm 1cm 1cm, clip, width=\textwidth]{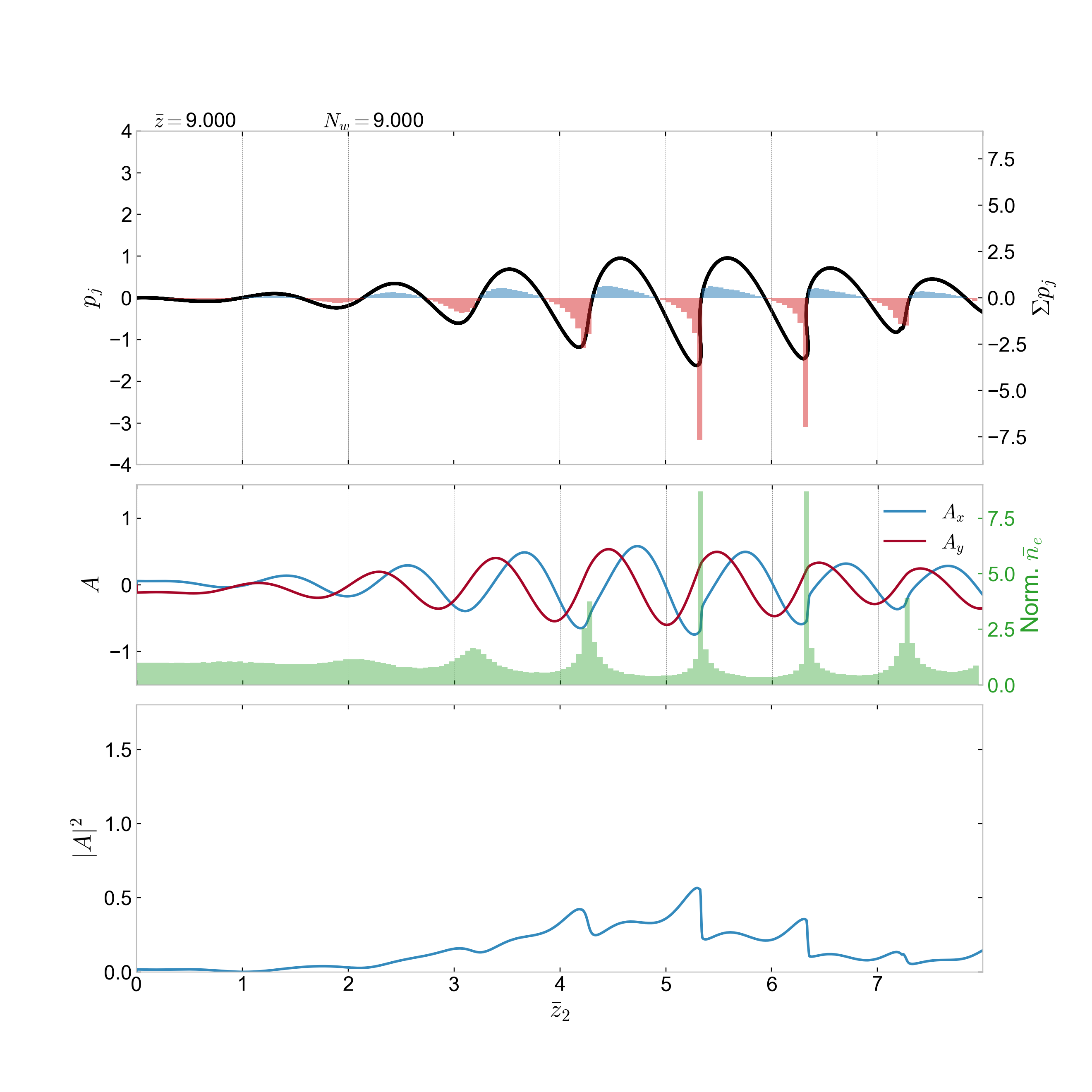}
        \caption{}
        \label{fig:zb9}
    \end{subfigure}
    \begin{subfigure}[c]{0.49\textwidth}
        \includegraphics[trim=1.5cm 1cm 1cm 1cm, clip, width=\textwidth]{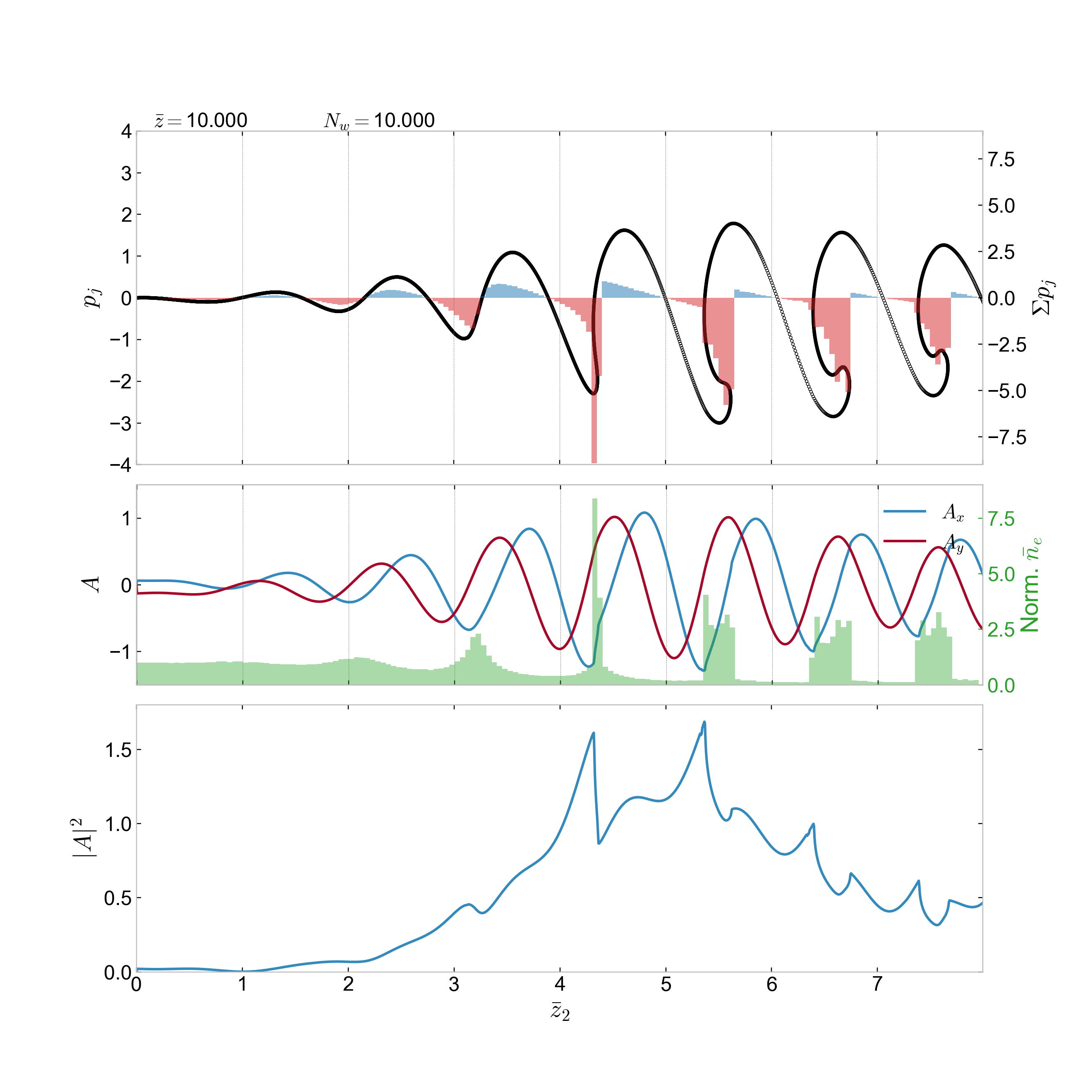}
        \caption{}
        \label{fig:zb10}
    \end{subfigure}

    \begin{subfigure}[c]{0.49\textwidth}
        \includegraphics[trim=0.5cm 0.5cm 0.1cm 0.1cm, clip, width=\textwidth]{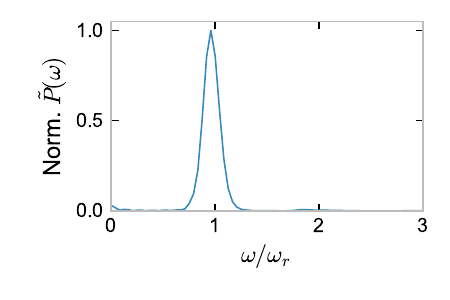}
        \caption{}
        \label{fig:psd_zb9}
    \end{subfigure}
    \begin{subfigure}[c]{0.49\textwidth}
        \includegraphics[trim=0.5cm 0.5cm 0.1cm 0.1cm, clip, width=\textwidth]{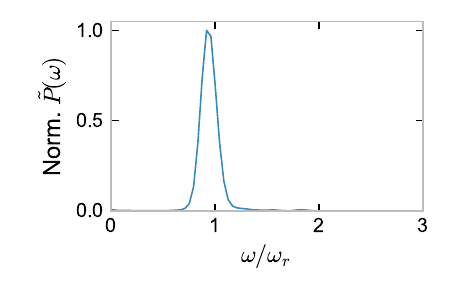}
        \caption{}
        \label{fig:psd_zb10}
    \end{subfigure}
    
    \caption{FEL electron phase-space and radiation evolution about post-linear evolution saturation of the radiation pulse. (a,c) are at $\bar{z}=9$, and (b,d) are at $\bar{z}=10$. Top (a,b):  The electron phase-space $(\bar{z}_{2_j}, p_j)$ (black dotted) and  localised net energy $\Sigma p_j$. Middle (a,b):  The radiation field components $A_x$, $A_y$ (blue and red solid lines) and  localised electron number density $\bar{n}_e$ (green bars) as  normalised to 1 for a `fresh' unbunched beam. The bar plots for the localised electron parameters $\Sigma p_j$ and $\bar{n}_e$ are within the bins of  width $\lambda_r/20$. Bottom (a,b): The scaled radiation power $|A(\bar{z}_2)|^2 $. Plots (c,d) are of the scaled spectral power $\tilde{P}(\omega)$ as a function of frequency scaled with respect to the resonant frequency $\omega_r$. Note that as $4\pi\rho = 1$ in this simulation, a resonant electron propagates one undulator period for a change of $\Delta \bar{z} = 1$, and one radiation period corresponds to a change of $\Delta \bar{z}_2 = 1$. {\bf Electronic version:} To see an animation of the interaction over 2 undulator periods click figure 2 a) or b). An animation of the full evolution from the start of the undulator to full saturation is also available from~\cite{video}.}
    \label{fig:zbar=9-10}
\end{figure*}

The results of the simulation up to a normal steady-state type FEL saturation process are shown in figure~\ref{fig:zbar=9-10} at undulator positions $\bar{z}=9$ and $10$ over the observation interval $0 \leq \bar{z}_2 \leq 8$ in the interaction window as shown in figure~\ref{fig:windows}. 

The electron phase-space $(\bar{z}_{2_j}, p_j)$, scaled circularly polarised radiation fields components $A_x, A_y$, the corresponding scaled intensity $|A|^2$ and the scaled power spectrum $\tilde{P}$ are plotted.
The localised electron number density $\bar{n}_e$, initially $\bar{n}_e=1$ for the `fresh' uniform electron beam at $\bar{z}=0$, and localised energy $\Sigma p_j$, are also plotted.

In figure~\ref{fig:zb9}, at $\bar{z} = 9$, the interaction is approaching the `normal', post-linear evolution FEL saturation process, as described  in~\cite{mcneil2010x}. The electrons are seen to be strongly bunched about the peak of the radiation pulse with a bunch spacing at the fundamental radiation wavelength $(\Delta \bar{z}_2 \sim 1)$. Notice that the bunched electrons drive the field locally at the sub-wavelength scale, unlike in averaged simulations.

In figure~\ref{fig:zb10}, at $\bar{z} = 10$, the electron bunching is seen to saturate and begin to de-bunch around the peak of the radiation power at $(\bar{z}_2 \sim 5.5)$. For $\bar{z}_2 > 5.5$ the de-bunching electrons are seen to start re-absorbing energy from the radiation causing its pulse duration to start to decrease post-saturation. At both values of $\bar{z}$, the power spectrum is is seen to be centred about resonance $\omega / \omega_r=1$. 

\begin{figure*}
    \centering
    \begin{subfigure}[c]{0.49\textwidth}
        \includegraphics[trim=1cm 1cm 1cm 1cm, clip, width=\textwidth]{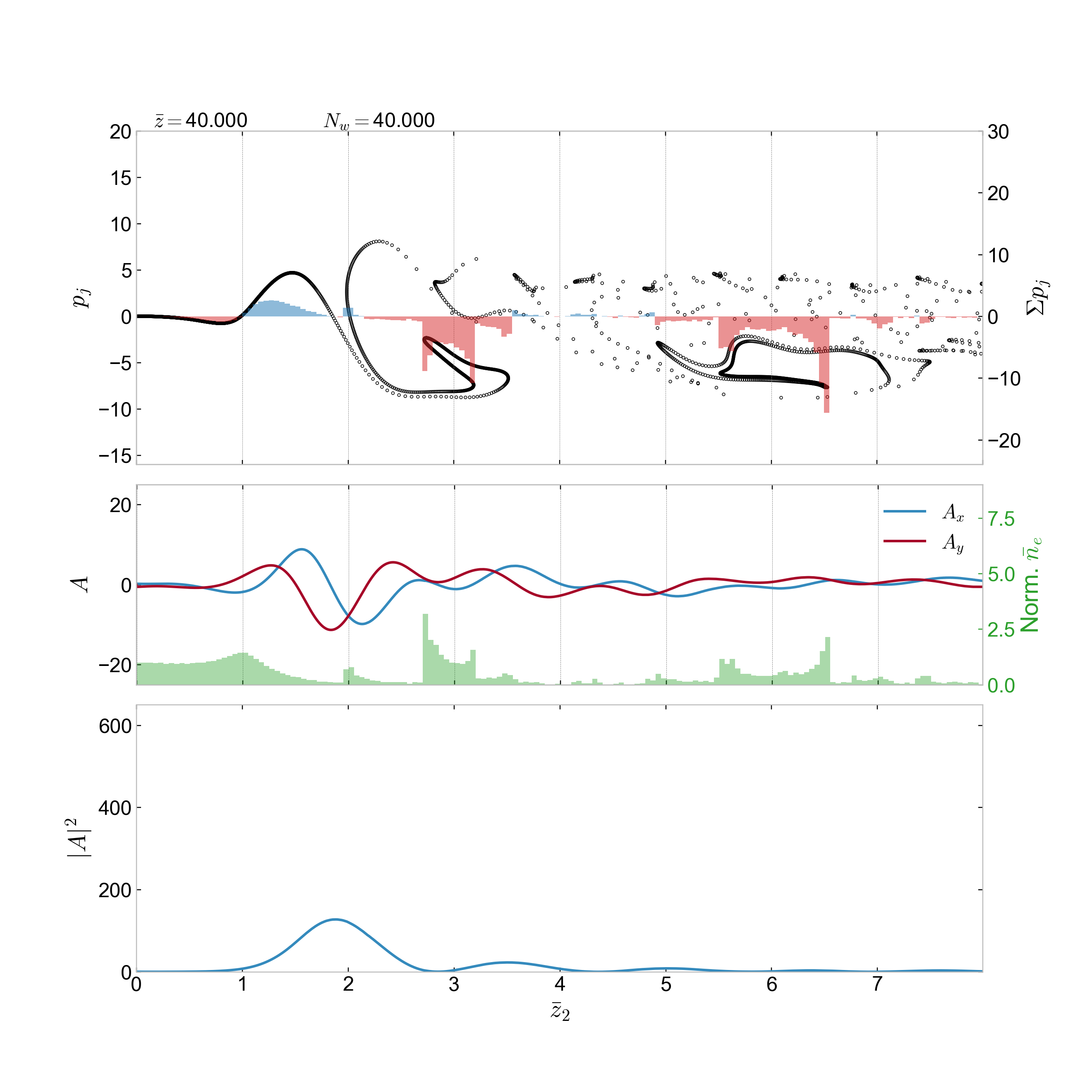}
        \caption{}
        \label{fig:zb40}
    \end{subfigure}
    \begin{subfigure}[c]{0.49\textwidth}
        \includegraphics[trim=1cm 1cm 1cm 1cm, clip, width=\textwidth]{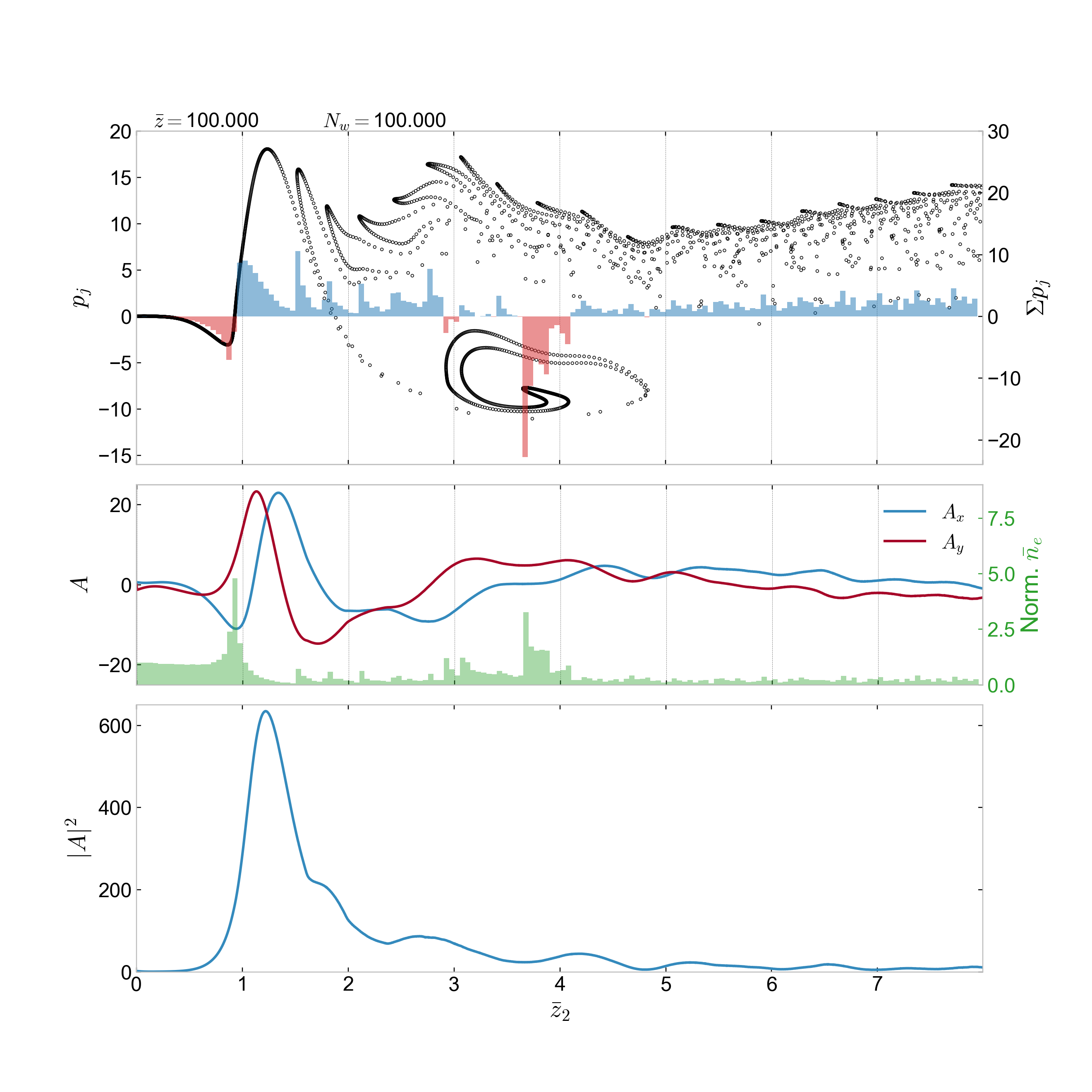}
        \caption{}
        \label{fig:zb100}
    \end{subfigure}
    \begin{subfigure}[c]{0.49\textwidth}
        \includegraphics[trim=0.5cm 0.5cm 0.1cm 0.1cm, clip, width=\textwidth]{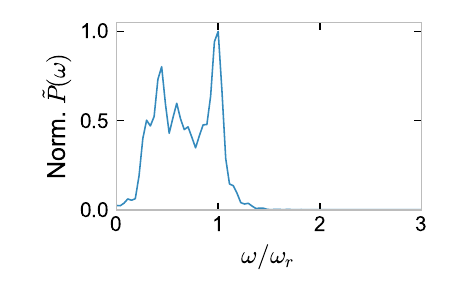}
        \caption{}
        \label{fig:psd_zb40}
    \end{subfigure}
    \begin{subfigure}[c]{0.49\textwidth}
        \includegraphics[trim=0.5cm 0.5cm 0.1cm 0.1cm, clip, width=\textwidth]{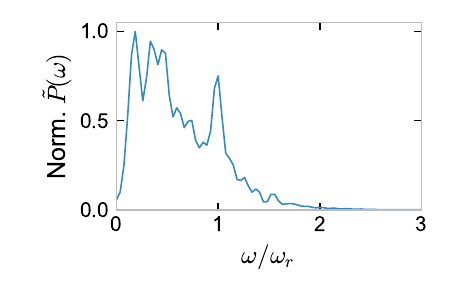}
        \caption{}
        \label{fig:psd_zb100}
    \end{subfigure}
    \caption{As figure \ref{fig:zbar=9-10}, but for scaled undulator distances of $\bar{z} = 40$ (left) and $\bar{z} = 100$ (right). {\bf Electronic version:} To see an animation of the interaction over 2 undulator periods click figure 3 a) or b).  An animation of the full evolution from the start of the undulator to full saturation is also available from~\cite{video}.}
    \label{fig:phase-space_power_PSD_zb40-100}
\end{figure*}

For further interaction to larger values of $\bar{z} > 10$, it can be expected that this process continues each successive undulator period, with maximum electron bunching and coherent emission occurring just before the peak of the radiation power and then de-bunching and re-absorbing. This then drives the growth of the scaled radiation peak power while reducing its pulse duration. This regime is shown in figure~\ref{fig:phase-space_power_PSD_zb40-100} where a breakdown of the averaged theory model is seen to occur, with electron bunching deviating from being spaced at the resonant wavelength and radiation powers approximately constant over a radiation wavelength. 

In figure~\ref{fig:zb40}, at the scaled undulator distance of $\bar{z} = 40$, the electron pulse is seen to enter the leading radiation pulse `edge' at $\bar{z}_2 \sim 1$, initiating the bunching process as seen from the phase-space and normalised electron density, $\bar{n}_e$ (middle).
Note that the electron density variations for $\bar{z}_2 > 2$  mainly originate from the interaction during previous undulator periods.
The electron bunching process is seen to occur within one radiation period over one undulator period positioned about the main radiation peak power about $\bar{z}_2 \sim 2$.

During one undulator period ($\Delta\bar{z} = 1$), the electrons strongly bunch within the short radiation pulse, centred at $\bar{z}_2\sim 2$ and of Full Width at Half Maximum (FWHM) in $\bar{z}_2$ of $\tau_p\approx 1$, lose energy  and then propagate out of the main pulse, as seen in the interval $2.5 < \bar{z}_2 < 3.5$ for $p_j<0$. Some electrons also gain energy to values of $p_j>0$. These electrons do not propagate as quickly to larger values of $\bar{z}_2$ as those that have lost energy with $p_j<0$. They also tend to retain their higher energy over many undulator periods. The energies gained are significantly greater than those during a normal FEL process where maximum gains are $p_j \sim 1$ - see figure~\ref{fig:zb9}. As may be expected and subsequently shown, the energy gain of those electrons increases as the spike power increases.

Note that some of the electrons, in the interval $2.7<\bar{z}_2 < 3.5$, have lost and then re-gained energy in their interaction with the radiation pulse. This is because of their initial large energy loss from $p_j =0$ to $p_j \sim -8$ which happens in less than one undulator period $(\bar{z}<1)$ as they enter the radiation pulse and encounter its rapidly increasing power. This energy loss causes them to propagate rapidly in $\bar{z}_2$ in the high power pulse within one undulator period, from a radiation emission phase, where they lose energy, to a radiation absorption phase, where they gain energy.

The electrons continue to emit and absorb radiation as they propagate away from the initial radiation pulse, but with reduced bunching, with a secondary pulse at $\bar{z}_2 \sim 3$ and further  pulses with reducing powers following.

The electrons traversing this pulse structure for $\bar{z}_2 > 3$ are seen to have bunching structures greater than one radiation wavelength for the electrons with lower lower energies, and less than one radiation wavelength for the electrons with higher energies. Those at the lower energies radiate at longer wavelengths  and those at higher energies emit at shorter wavelengths. This can be seen in the associated scaled radiation  power spectrum in figure~\ref{fig:psd_zb40}, with a prominent lower frequency peak $(\omega/\omega_r \sim 0.5)$. The electrons that have increased in energy, with a (smaller) sub-wavelength bunching structure contribute to an increased emission at frequencies just above resonance $(\omega/\omega_r \sim 1)$. This non-linear radiation pulse emission still maintains similarities to the superradiant structures as derived in~\cite{piovella1991hyperbolic,zen2023full, piovella2021review}.

In figure~\ref{fig:zb100}, at the scaled undulator distance of $\bar{z} = 100$, it can be seen that process described for figure~\ref{fig:zb40} has continued, with the pulse peak radiation power increasing and its duration reducing. This increased pulse power and reduced duration reduces the time within an undulator period that electrons experience within the first peak, now centred at $\bar{z}_2 \sim 1.2$. This leads to the electrons entering the subsequent radiation pulse(s), centered at larger $\bar{z}_2$, within one undulator period and leads to a more complex main-pulse/sub-pulse interaction with the electrons.
Note that the pulse width of the main peak has now reduced to just over half of the radiation wavelength, $\tau_p \sim 0.5$. 

For both $\bar{z} = 40$ and $100$, the distance between the lower energy electron bunches that have interacted with the radiation pulse (and subsequently propagated outside of the observation window in the case of figure~\ref{fig:psd_zb100}) exceeds two fundamental wavelengths. This is seen to result in significant enhanced emission at lower frequencies, $\omega/\omega_r < 0.5$, as seen in the spectral power of figures~(\ref{fig:psd_zb40}, \ref{fig:psd_zb100}).

In figure~\ref{fig:psd_zb100}, a further characteristic in this regime is the rapid energy fluctuation of electrons that initialy lose energy on entering the pulse. This is due, in part, to the fast inter-wavelength scale movement of the electrons within one undulator period. Those electrons that initially gain energy from the radiation pulse can achieve a relatively high energy value of $p_j \sim 10$ within the first peak. These accelerated electrons form a more stable, shorter period, electron bunching band after the first peak at $\bar{z}_2 > 2$.  This behaviour is evident in the $\Sigma p_j$ and $\bar{n}_e$ plots of figure~\ref{fig:zb100}(top and middle). These smaller, higher energy, electron bunches, positioned within the interval of $1 < \bar{z}_2 < 3$, have spacings that are less than the fundamental radiation wavelength, and consequently there is a noticeable broadening of the bandwidth in the higher frequency range as seen in figure~\ref{fig:phase-space_power_PSD_zb40-100}d, for $\omega/\omega_r >1$.

\begin{figure*}
    \centering
    \begin{subfigure}[c]{0.49\textwidth}
        \includegraphics[trim=1cm 1cm 1cm 1cm, clip, width=\textwidth]{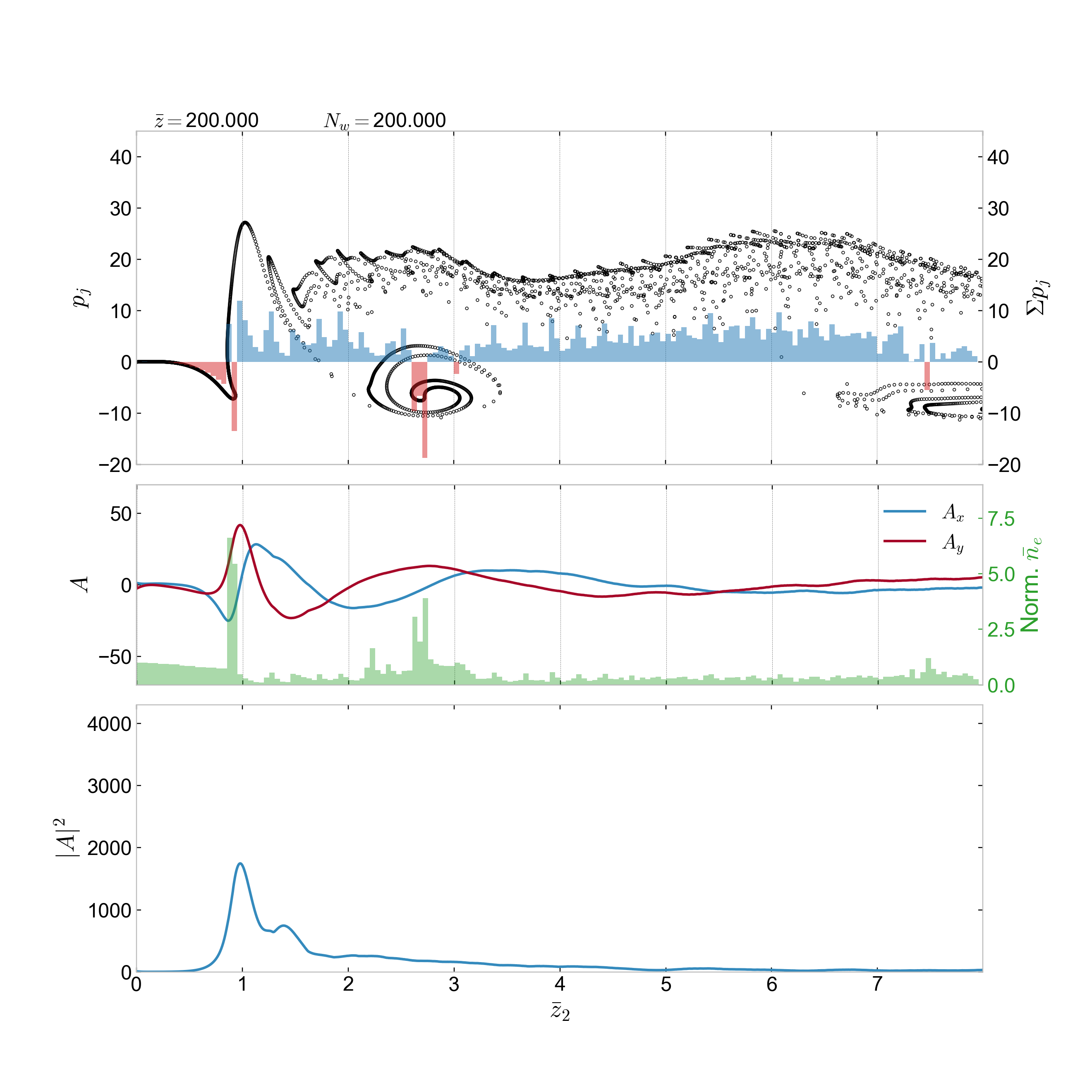}
        \caption{}
        \label{fig:zb200}
    \end{subfigure}
    \begin{subfigure}[c]{0.49\textwidth}
        \includegraphics[trim=1cm 1cm 1cm 1cm, clip, width=\textwidth]{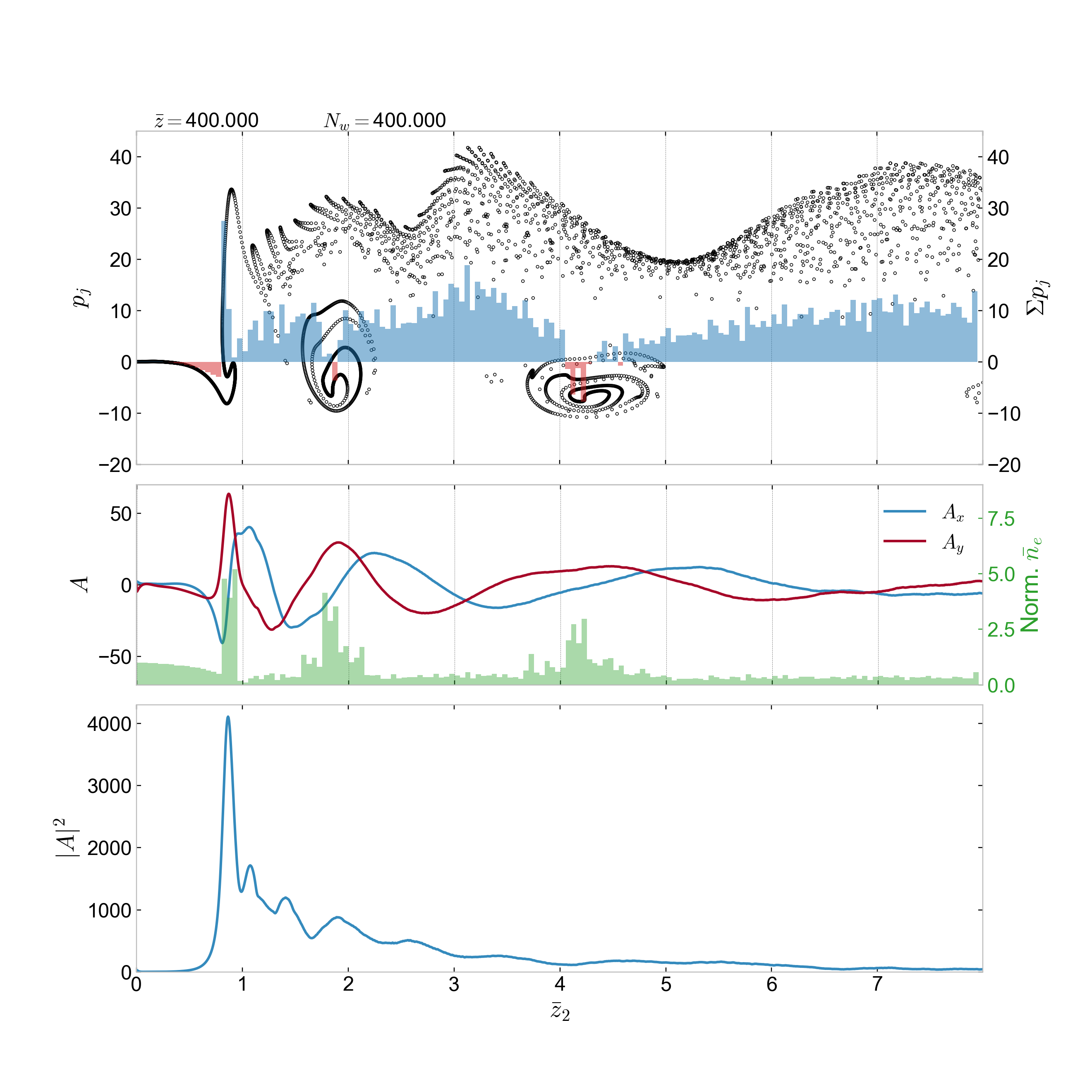}
        \caption{}
        \label{fig:zb400}
    \end{subfigure}

    \begin{subfigure}[c]{0.49\textwidth}
        \includegraphics[trim=0.5cm 0.5cm 0.1cm 0.1cm, clip, width=\textwidth]{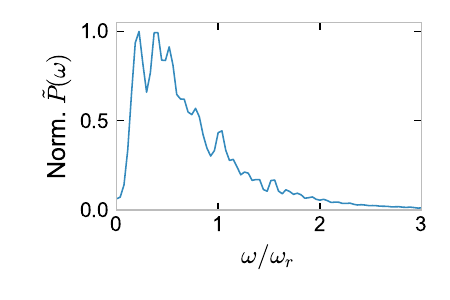}
        \caption{}
        \label{fig:psd_zb200}
    \end{subfigure}
    \begin{subfigure}[c]{0.49\textwidth}
        \includegraphics[trim=0.5cm 0.5cm 0.1cm 0.1cm, clip, width=\textwidth]{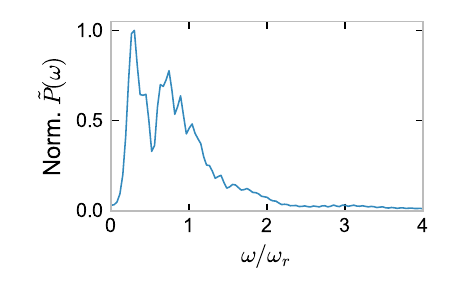}
        \caption{}
        \label{fig:psd_zb400}
    \end{subfigure}  
    \caption{As figure \ref{fig:zbar=9-10}, but for scaled undulator distances of $\bar{z} = 200$ (left) and $\bar{z} = 400$ (right). {\bf Electronic version:} To see an animation of the interaction over 2 undulator periods click figure 4 a) or b).  An animation of the full evolution from the start of the undulator to full saturation is also available from~\cite{video}.}.
    \label{fig:phase-space_power_PSD_zb200-400}
\end{figure*}

In figure~\ref{fig:phase-space_power_PSD_zb200-400}, the interaction has progressed to $\bar{z}=200$ (left) and $\bar{z}=400$ (right). The first radiation peak about $\bar{z}_2\approx 1$ is seen to have further increased in power, and with a reduced width to significantly less than one radiation wavelength. While the electrons that gain energy have increased to higher values, it is seen that those that lose energy attain similar values to those of figure~\ref{fig:phase-space_power_PSD_zb40-100}. 
This increased power and reduced width of the radiation pulse induces the electrons to exhibit significant energy fluctuations in a short interval of propagation through the pulse. Electrons can then propagate through the pulse of width $\tau_p < 0.25$ in less than a quarter of an undulator period, experiencing both energy loss and gain as they do so. This is evidenced by the greater spiraling in phase space of the lower energy electrons that have propagated through the initial pulse when compared with those of figure~\ref{fig:phase-space_power_PSD_zb40-100}.

As $\bar{z}$ increases, and the peak pulse decreases in width, higher frequency components of the field with $\omega/\omega_r>1$ receive further growth. Electrons that have left the radiation peak, form bunches with a greater spacing in $\bar{z}_2 >1$, and consequently emit at a lower-frequency. This is seen in the lower frequency radiation fields on the trailing edge of the pulse, where a correlation between the electron bunches and the scaled field phase is noticeable. This correlation is seen to continue to larger $\bar{z}$ as the interaction progresses.

\begin{figure*}
    \centering
    \begin{subfigure}[c]{0.49\textwidth}
        \includegraphics[trim=1cm 1cm 1cm 1cm, clip, width=\textwidth]{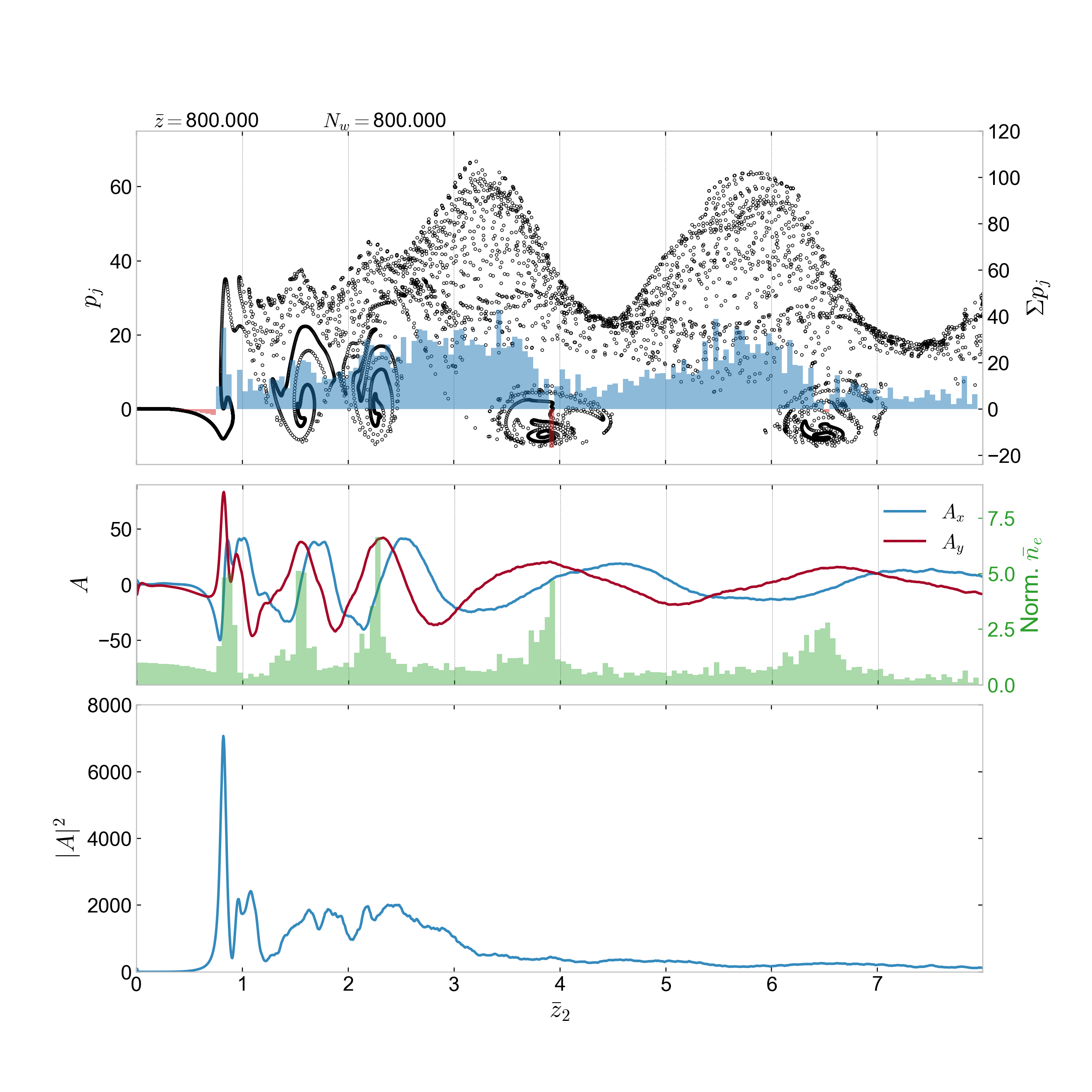}
        \caption{}
        \label{fig:zb800}
    \end{subfigure}
    \begin{subfigure}[c]{0.49\textwidth}
        \includegraphics[trim=1cm 1cm 1cm 1cm, clip, width=\textwidth]{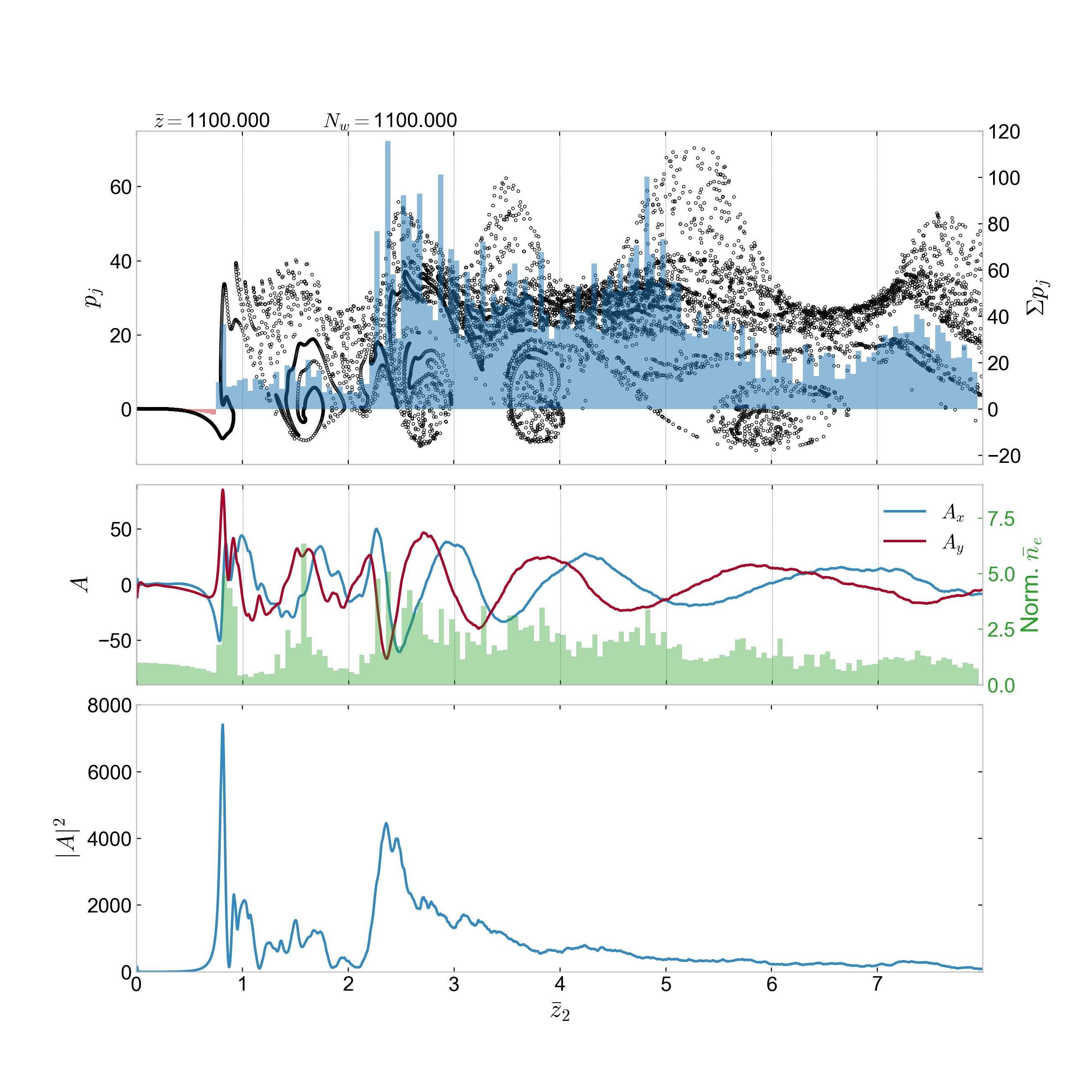}
        \caption{}
        \label{fig:zb1100}
    \end{subfigure}

    \begin{subfigure}[c]{0.49\textwidth}
        \includegraphics[trim=0.5cm 0.5cm 0.1cm 0.1cm, clip, width=\textwidth]{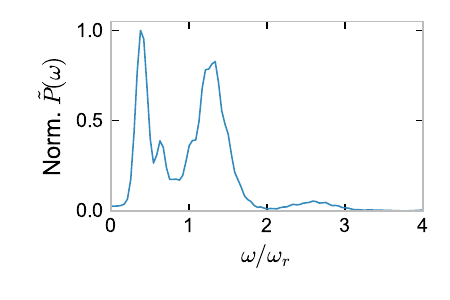}
        \caption{}
        \label{fig:psd_zb800}
    \end{subfigure}
    \begin{subfigure}[c]{0.49\textwidth}
        \includegraphics[trim=0.5cm 0.5cm 0.1cm 0.1cm, clip, width=\textwidth]{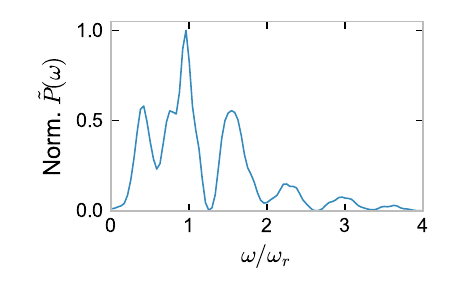}
        \caption{}
        \label{fig:psd_zb1100}
    \end{subfigure}  
    
    \caption{As figure \ref{fig:zbar=9-10}, but for scaled undulator distances of $\bar{z} = 800$ (left) and $\bar{z} = 1100$ (right). {\bf Electronic version:} To see an animation of the interaction over 2 undulator periods click figure 5 a) or b). An animation of the full evolution from the start of the undulator to full saturation is also available from~\cite{video}.}
    \label{fig:phase-space_power_PSD_zb800-1100}
\end{figure*}

In figure~\ref{fig:phase-space_power_PSD_zb800-1100}, the short, high power pulse appears to have entered into a highly non-linear saturated regime, where the growth of the pulse peak power and its width are very similar at both $\bar{z}=800$ (left) and $\bar{z}=1100$ (right). This pulse saturation process and scaling as a function of the FEL parameter $\rho$, are now investigated. 

\subsection{Pulse saturation}

As the interaction progresses to larger values of $\bar{z}$, the growth of the first radiation spike in the scaled radiation power is seen to have saturated with $|A_p|^2\approx 7000$ around $\bar{z}_2 \approx 0.8$, as seen in figure~\ref{fig:phase-space_power_PSD_zb800-1100}.
While in  figure~\ref{fig:zbar=9-10}a) it is seen that the average energy lost by the electrons immediately after passing through the first spike is large, causing the electron bunches to rapidly propagate in $\bar{z}_2$ so increasing the distance between bunches after the spike, in figure~\ref{fig:phase-space_power_PSD_zb800-1100}a) it is seen that the average energy lost reduces following saturation, causing the electrons to propagate more slowly in $\bar{z}_2$. This results in the electron bunches that have propagated through the first spike having a smaller separation in $\bar{z}_2$. Where the bunched electron separation is larger, it is seen that this contributes to longer wavelength emission. Note also that as saturation occurs, there are more electrons gaining energy from the spike. 

From a simplistic perspective, it is hypothesised that saturation of the first radiation spike occurs when the energy loss of electrons propagating through it means they can propagate a relatively large distance with respect to a radiation wavelength within one undulator period. From a scaling perspective, saturation is therefore defined to occur when an electron loses sufficient energy within one half of an undulator period  to propagate an additional one half of a resonant radiation wavelength~$\Delta \bar{z}_{2j}$ above its resonant slippage rate in the radiation frame of reference of $d\bar{z}_{2j}/d\bar{z}=1$, i.e.~$\Delta \bar{z}_{2j}=\Delta \bar{z}=2\pi\rho$. A simple estimate of the scaling using this definition is given in \ref{appendix}, from which the peak scaled power of the spike is given by $|A_p|^2 \approx 1/\pi^2\rho^4$ and its scaled energy is  $\varepsilon_p \approx\tau_p |A_p|^2 \approx  4f/\pi\rho^3$, where $\tau_p=f\times 4\pi \rho$ describes the spike width in $\bar{z}_2$ and $f$ is the spike width as a fraction of one radiation wavelength.

\begin{figure*}
    \centering
    \includegraphics[width=0.92\textwidth]{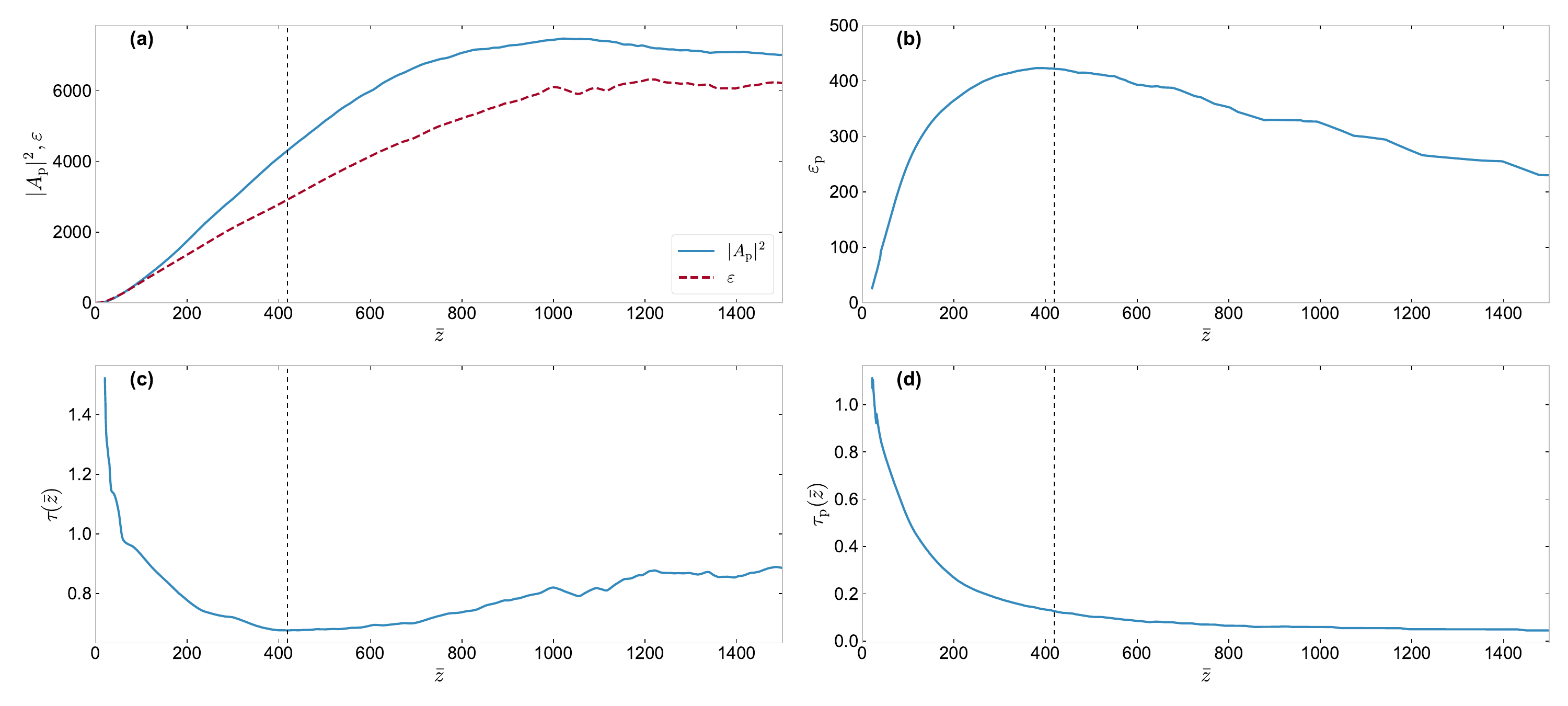}
    \caption{The FEL evolution as measured in the interval $0 \leq \bar{z}_2 \leq 9$ as a function of $\bar{z}$ for the case of $4\pi\rho = 1$. (a) scaled peak power $|A_p|^2$ and pulse energy $\varepsilon$; (b) scaled energy within the first peak (spike) $\varepsilon_{\text{p}}$; (c) the scaled pulse duration $\tau$ - the vertical dashed line shows the minimum defined as the point of saturation; and (d) the first radiation pulse (spike) duration $\tau_p$.}
    \label{fig:4pirho=1}
\end{figure*}

From the computer simulations, saturation is defined to be when the full pulse duration $\tau(\bar{z})$ over the simulation interval $0<\bar{z}_2<8$ is a minimum. This may be calculated from $\tau(\bar{z}) = \varepsilon(\bar{z})/|A_p(\bar{z})|^2$ where the scaled pulse energy over the simulation interval is $\varepsilon(\bar{z}) = \int |A(\bar{z},\bar{z}_2)|^2 d\bar{z}_2$ and, as above, the peak scaled power over the interval is given by $|A_p(\bar{z})|^2$. 
Simulation results as a function of $\bar{z}$ and for $4\pi\rho=1$, are shown in figure~\ref{fig:4pirho=1} where saturation is seen to occur at $\bar{z} \approx 419$, as highlighted  by the vertical dashed line in figure~\ref{fig:4pirho=1}c). While the peak power $|A_p(\bar{z})|^2$ and pulse energy $\varepsilon(\bar{z})$ continue to increase post-saturation, as seen in figure~\ref{fig:4pirho=1}a), the energy within the first peak $\varepsilon_{\text{p}}(\bar{z})$ (that of the radiation spike), is also seen to saturate at $\bar{z} \approx 419$ and then decreases with increasing $\bar{z}$ as shown in figure~\ref{fig:4pirho=1}b). 
While at  saturation the pulse duration $\tau(\bar{z})$ minimises, and then begins to increase with $\bar{z}$ (figure~\ref{fig:4pirho=1}c), the width of the first radiation spike, $\tau_p(\bar{z}) = \varepsilon_p(\bar{z})/|A_p(\bar{z})|^2$, is seen to continue to reduce as seen in figure~\ref{fig:4pirho=1}d).

\begin{figure}[ht]
    \centering
    \includegraphics[width=0.45\textwidth]{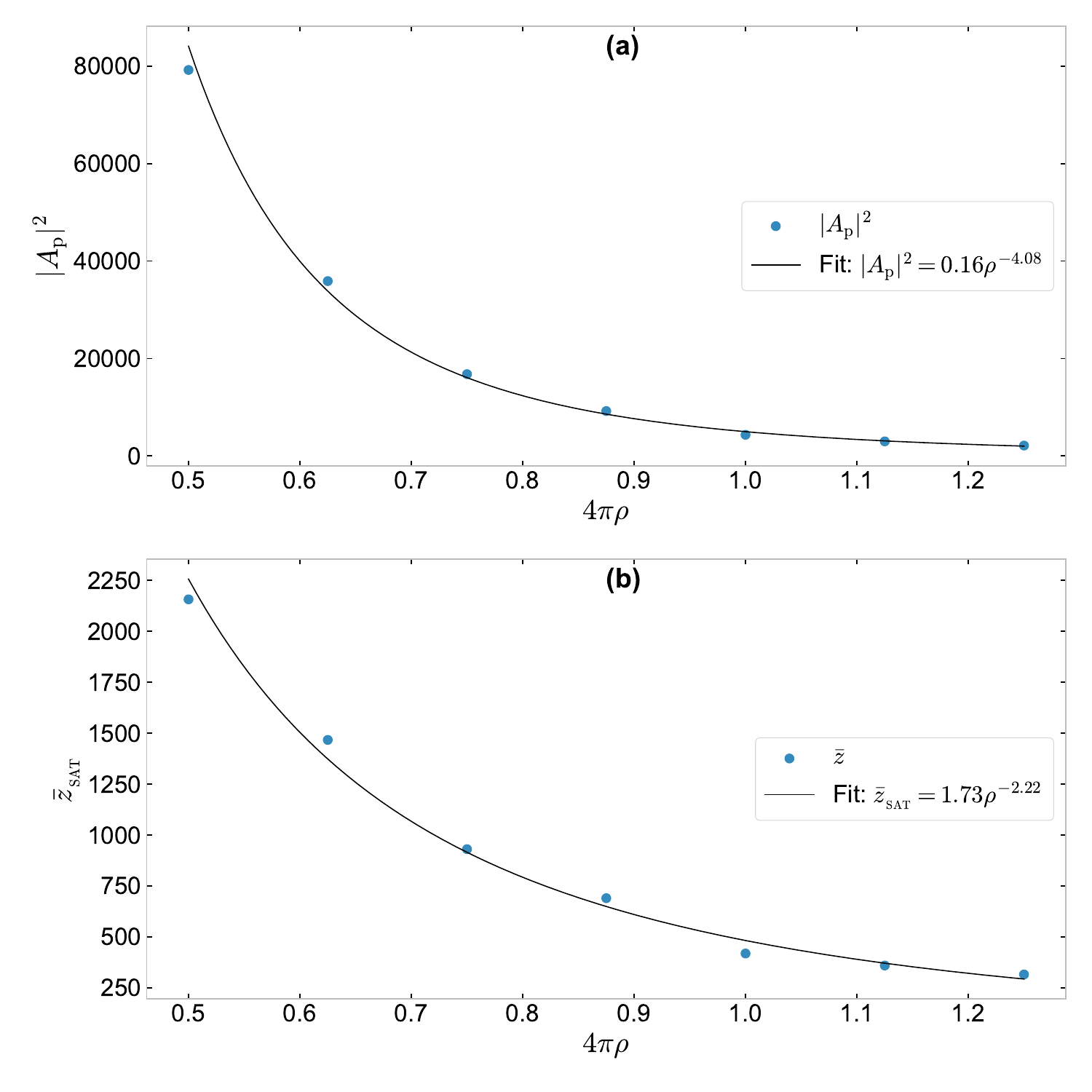}
    \caption{The saturated values of the simulations (dots) for (a) scaled peak power $|A_p|^2$, and (b) scaled saturation undulator length $\bar{z}_{_\text{sat}}$, as a function of $4\pi\rho$. The solid lines are the fitting functions as given in the box.}
    \label{fig:fit}
\end{figure}

\begin{figure}[ht]
    \centering
    \includegraphics[width=0.45\textwidth]{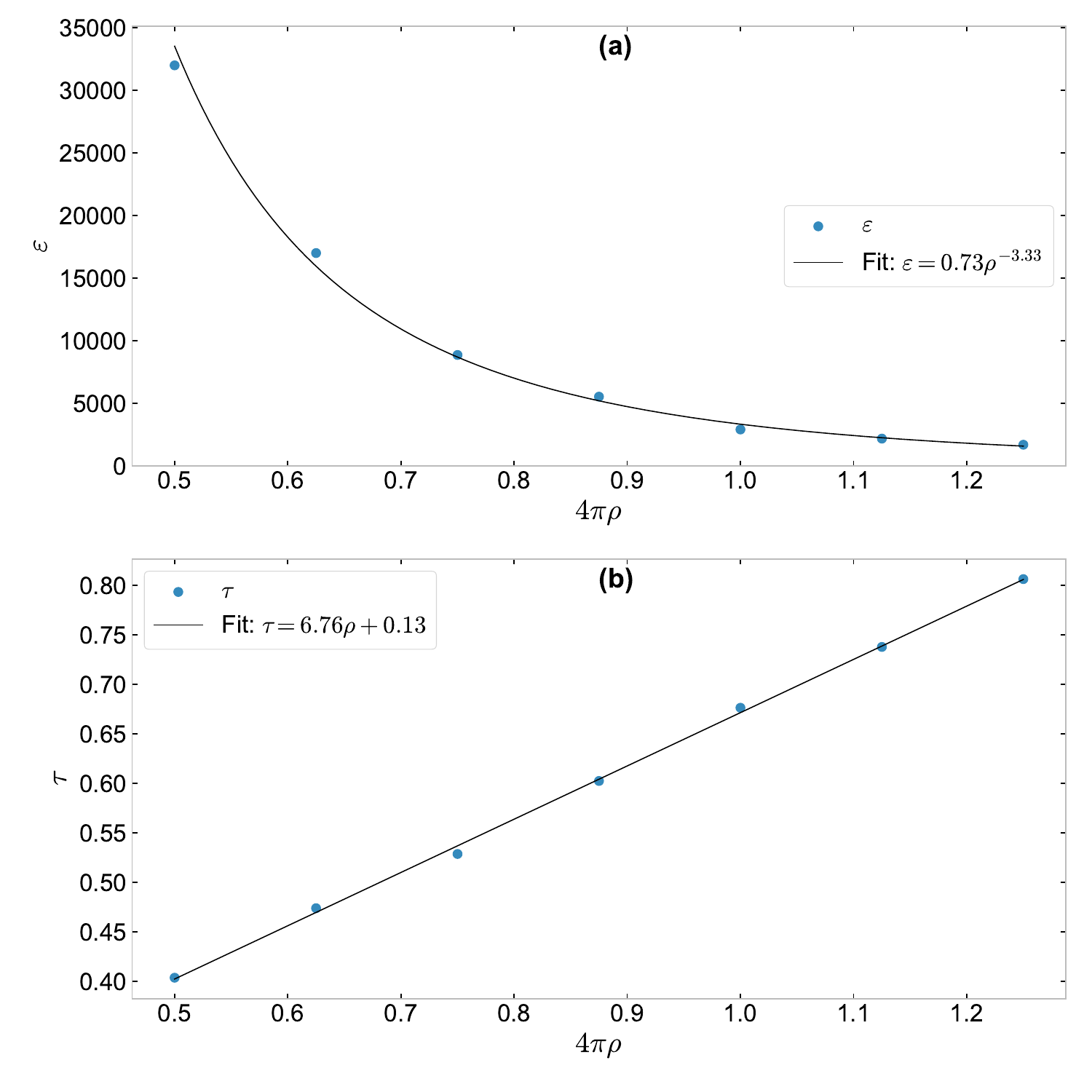}
    \caption{The saturated values of the simulations (dots) for (a) scaled pulse energy $\varepsilon$, and (b) pulse duration $\tau$, as a function of $4\pi\rho$. The solid lines are the fitting functions as given in the box.}
    \label{fig:fit2}
\end{figure}

\begin{figure}[ht]
    \centering
    \includegraphics[width=0.45\textwidth]{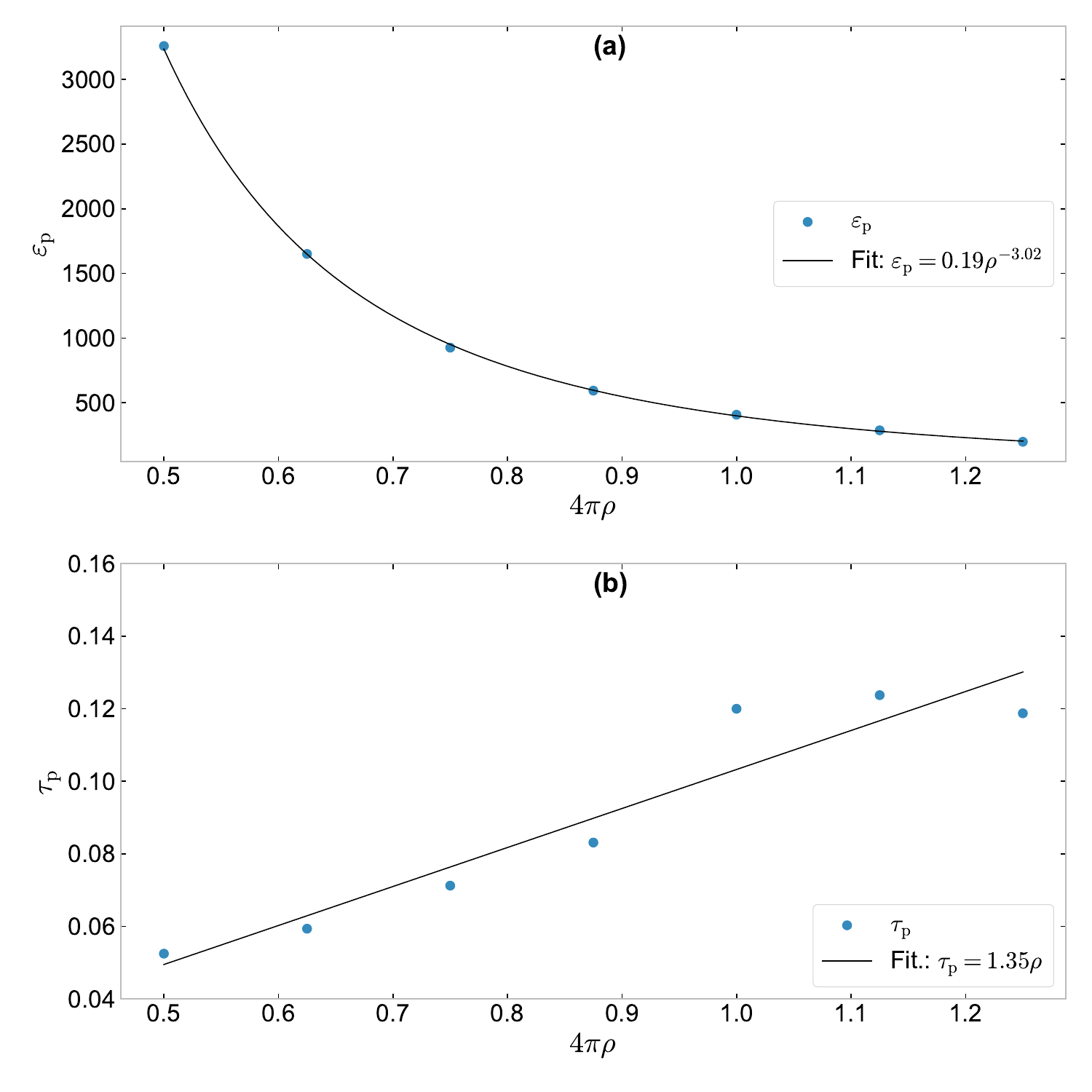}
    \caption{The saturated values of the simulations (dots) for (a) scaled energy of the first radiation pulse (spike) $\varepsilon_{\text{p}}$, and (b) its width $\tau_p$, as a function of $4\pi\rho$. The solid lines are the fitting functions as given in the box.}
    \label{fig:fit3}
\end{figure}

The simulation of above was also carried out for a range of values of $4 \pi \rho = 0.5\ldots 1.25$. 
(For $\gamma_r = 100$, these values of $\rho$ typical span an FEL operational wavelengths from the Far-Infrared to THz range.)
This allows comparison of the saturated values, when $\tau(\bar{z})$ is a minimum,  with  the simple scaling analysis of~\ref{appendix}. 
These  saturated values, together with their best-fit scalings,  are plotted in figure~\ref{fig:fit}, for the peak power of the radiation $|A_p|^2$ and the saturation undulator length $\bar{z}_{\text{sat}}$; in figure~\ref{fig:fit2}, for the radiation pulse energy $\varepsilon$ and pulse duration $\tau$ over the simulation interval; and in figure~\ref{fig:fit3} for the pulse energy $\varepsilon_p$ and duration $\tau_p$ of the first radiation spike. 

Using the above definition of saturation, and the simple scaling analysis of~\ref{appendix}, estimates of the scaled peak radiation power and energy of the radiation spike were obtained from~\ref{peakPower} and~\ref{peakEnergy} respectively. For example, in the simulations with $4\pi\rho =1$ above, the estimated saturation values are $|A_p|^2\approx 2500$ and $\varepsilon_p\approx 300$ for the fractional factor $f=0.12$ as taken from figure~\ref{fig:fit3}b). Given the significant approximations used in~\ref{appendix}, the relatively good agreement with the (highly non-linear) computational results, of figure~\ref{fig:fit}a) for $|A_p|^2$ and figure~\ref{fig:fit3}a) for $\varepsilon_p$, appears to be good, giving credence to the above hypothesis of the radiation spike saturation process.
This is further demonstrated with good agreement between the analysis and best-fit scaling of the simulations for  both $|A_p|^2 \propto \rho^{-4}$ (figure~\ref{fig:fit}a and~\ref{peakPower}) and $\varepsilon_p\propto \rho^{-3}$ (figure~\ref{fig:fit3}a and~\ref{peakEnergy}).


It is evident from figure~\ref{fig:fit}a, \ref{fig:fit2}a, and \ref{fig:fit3}a that as $\rho$ decreases greater radiation peak powers and pulse energies are required  to achieve saturation. This is because the lower $\rho$ values cause a weaker interaction between the electrons and the radiation field, so demanding an increased radiation power to attain the same level of energy exchange required to achieve saturation. 
In addition,  a shorter scaled pulse duration is required, as seen from figure~\ref{fig:fit2}b. The first peak duration $\tau_p$, expressed in units of radiation wavelength $\lambda_r$, remains approximately constant for all values of $4\pi\rho$, as shown in figure~\ref{fig:fit3}b.
This strongly suggests a single-cycle type limit where the radiation field assumes its minimal realistic duration. The use of an longer undulator to attain saturation is also seen in figure~\ref{fig:fit}b.
Consequently, modelling such saturation behaviour at smaller values of $\rho$ is  increasingly computationally demanding.

A video of the full evolution of the superradiant spike from the start of the undulator to full saturation is available from~\cite{video}.

\subsection{Estimation of diffractive effects}

The simulation and scaling results presented above are in one-dimension (1D), and so  exclude any three-dimensional (3D) effects such as radiation diffraction. It can be expected that the longer wavelengths of the radiation in the above simulations would diffract away from the electron beam, assumed to have a constant radius $\text{w}_0$. Assuming there are no  optical or gain `guiding' effects~\cite{mcneil2010x} acting on the radiation by the electron beam, then the radiation will diffract in accordance with its Rayleigh range, $Z_R = \pi \text{w}_0^2/\lambda\propto \omega$.
This will result in an increased decoupling between the electron beam and its radiation output at longer wavelengths/lower frequencies as the interaction progresses through the undulator. 

Assuming a transverse Gaussian radiation profile, the intensity $I$ can be expressed as a function of its peak power, $P_0$ on propagating  a distance $z$ through the undulator as:
\begin{equation}
    I(z) = \frac{2 P_0}{\pi \text{w}^2(z)},
\label{Idef}
\end{equation}
where $w(z)$ is the transverse radiation beam size along the $z$ axis:
\begin{equation}
    \text{w}(z) = \text{w}_0 \sqrt{1 + \left(\frac{z}{Z_R}\right)^2}.
\end{equation}

For increasingly larger distances $z>Z_R$, $\text{w}(z)$  tends to become inversely proportional to $Z_R$, and so to $\omega$. Hence, the radiation intensity $I(z)$ of~(\ref{Idef}) and its corresponding power $P(z)$, emitted by and then co-propagating a distance $z$ with the electron beam, tend to become directly proportional to the frequency as $I(z) \propto \omega^2$. 

This approximation is now used to estimate the effects of diffraction on the Power Spectral Density (PSD) of the radiation emission and its subsequent effect on the temporal pulse output. While this method is relatively simplistic, and cannot be expected to provide an accurate  model of any full 3D simulation of a the FEL interaction, it does provide a first estimate of what may be expected.

The frequency spectrum of the scaled electric field $A(\bar{z}_2)$ at a given point in the undulator is defined as $\tilde{A}(\bar{\omega}) = |\mathcal{F}[A(\bar{z}_2)]$ where $\mathcal{F}$ is the Fourier transform in $\bar{z}_2$ with respect to $\bar{\omega}=\omega/\omega_r$, and with the scaled PSD given by $\tilde{P}(\bar{\omega}) = |\tilde{A}(\bar{\omega})|^2$.
The scaled PSD of the radiation including diffractive effects due to propagation  scales with $\bar{\omega}^2$ as above in equation~(\ref{Idef}), and the spectrum, including 3D diffraction, is then estimated as $\tilde{P}_{3D}(\bar{\omega}) = |\tilde{A}_{3D}(\bar{\omega})|^2= |\bar{\omega}\tilde{A}(\bar{\omega})|^2$. The scaled intensity can then be retrieved via the inverse Fourier transform $|A_{3D}(\bar{z}_2)|^2 = |\mathcal{F}^{-1}[\tilde{A}_{3D}(\bar{\omega})]|^2$. 

\begin{figure*}[hbp]
    \centering
    \centering
    \begin{subfigure}[c]{0.49\textwidth}
        \includegraphics[width=\textwidth]{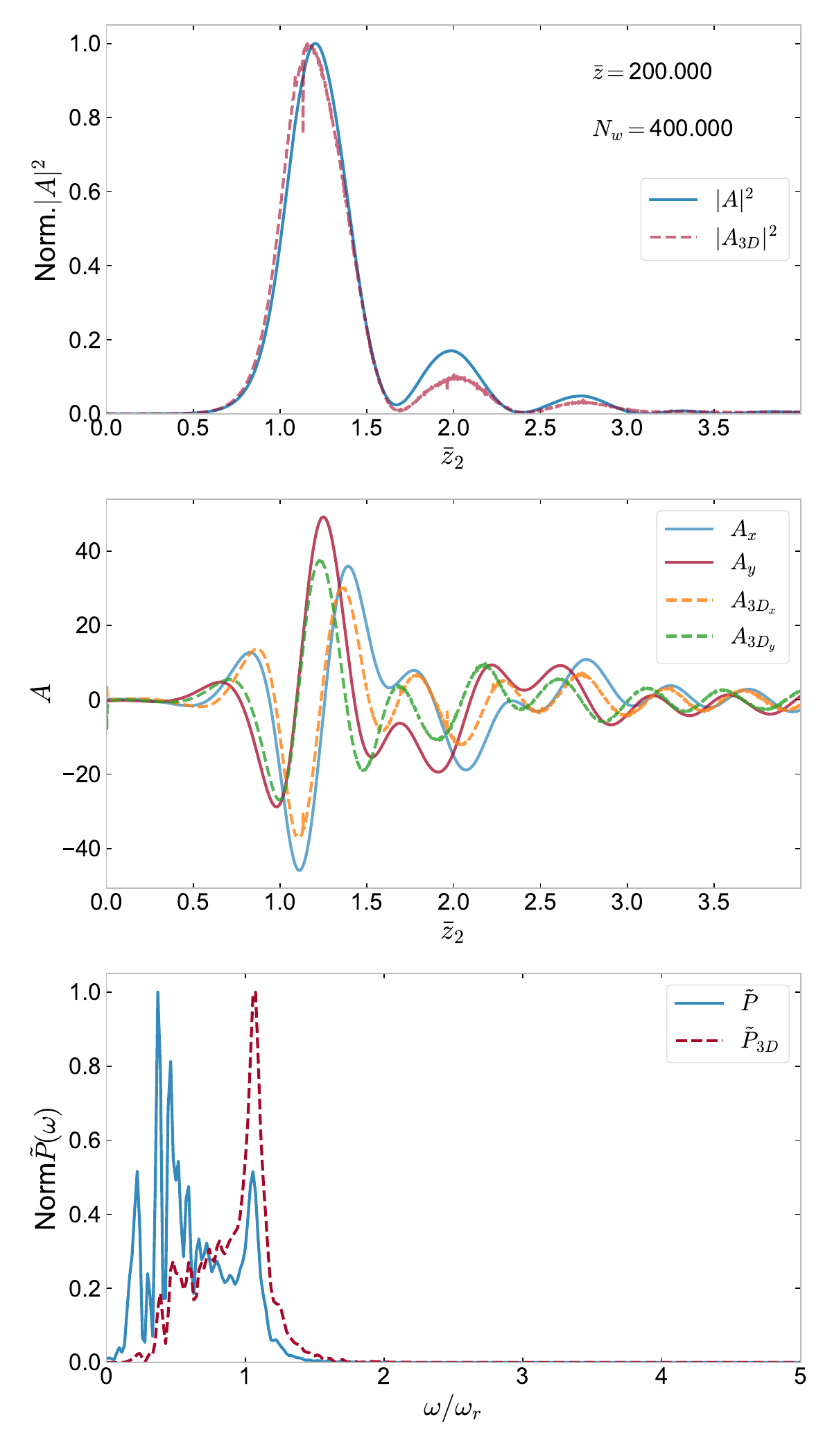}
        \caption{}
        \label{fig:PSD_4pirho=0.5_zbar=200}
    \end{subfigure}
    \begin{subfigure}[c]{0.49\textwidth}
        \includegraphics[width=\textwidth]{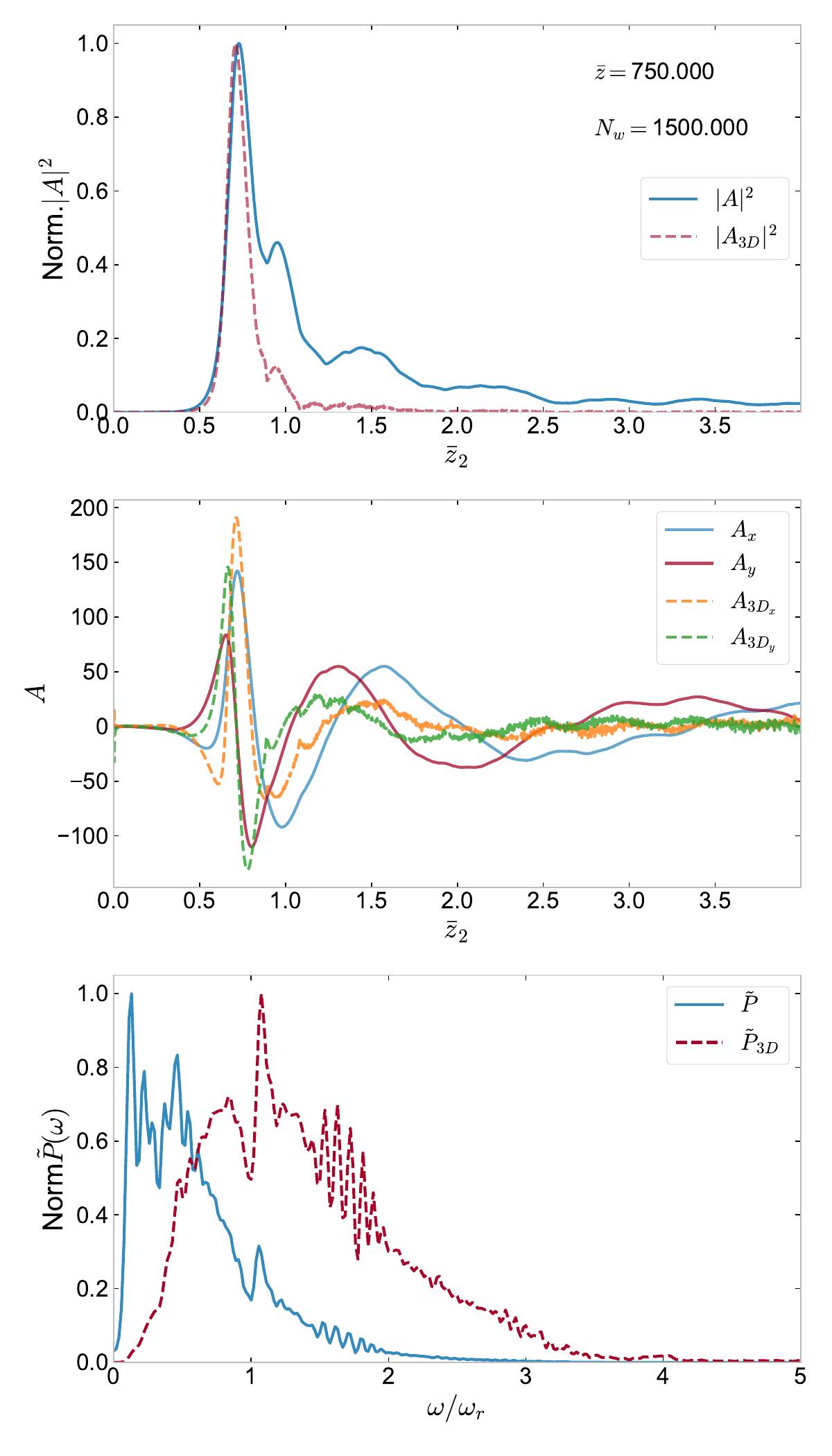}
        \caption{}
        \label{fig:PSD_4pirho=0.5_zbar=750}
    \end{subfigure}
    \caption{Comparison of 1D simulations with the 3D approximation for diffractive effects in the post-saturated superradiant regime for (a) $\bar{z} = 200$ and (b) $\bar{z} = 750$. (Top) Normalised 1D scaled power $|A|^2$ (solid blue line) and the 3D diffraction approximation $|A_{3D}|^2$ (dashed red line). (Middle) Scaled 1D electric field components $A_{x,y}$ (solid line) and the 3D approximation $A_{3D_{x,y}}$(dashed line). (Bottom) The 1D scaled Power Spectral Density $\tilde{P}$ (solid line) and the 3D diffraction approximation $\tilde{P}_{3D}$ (dashed line).}
    \label{fig:PSD_4pirho=0.5}
\end{figure*}

\begin{figure*}
    \centering
    \centering
    \begin{subfigure}[c]{0.49\textwidth}
        \includegraphics[width=\textwidth]{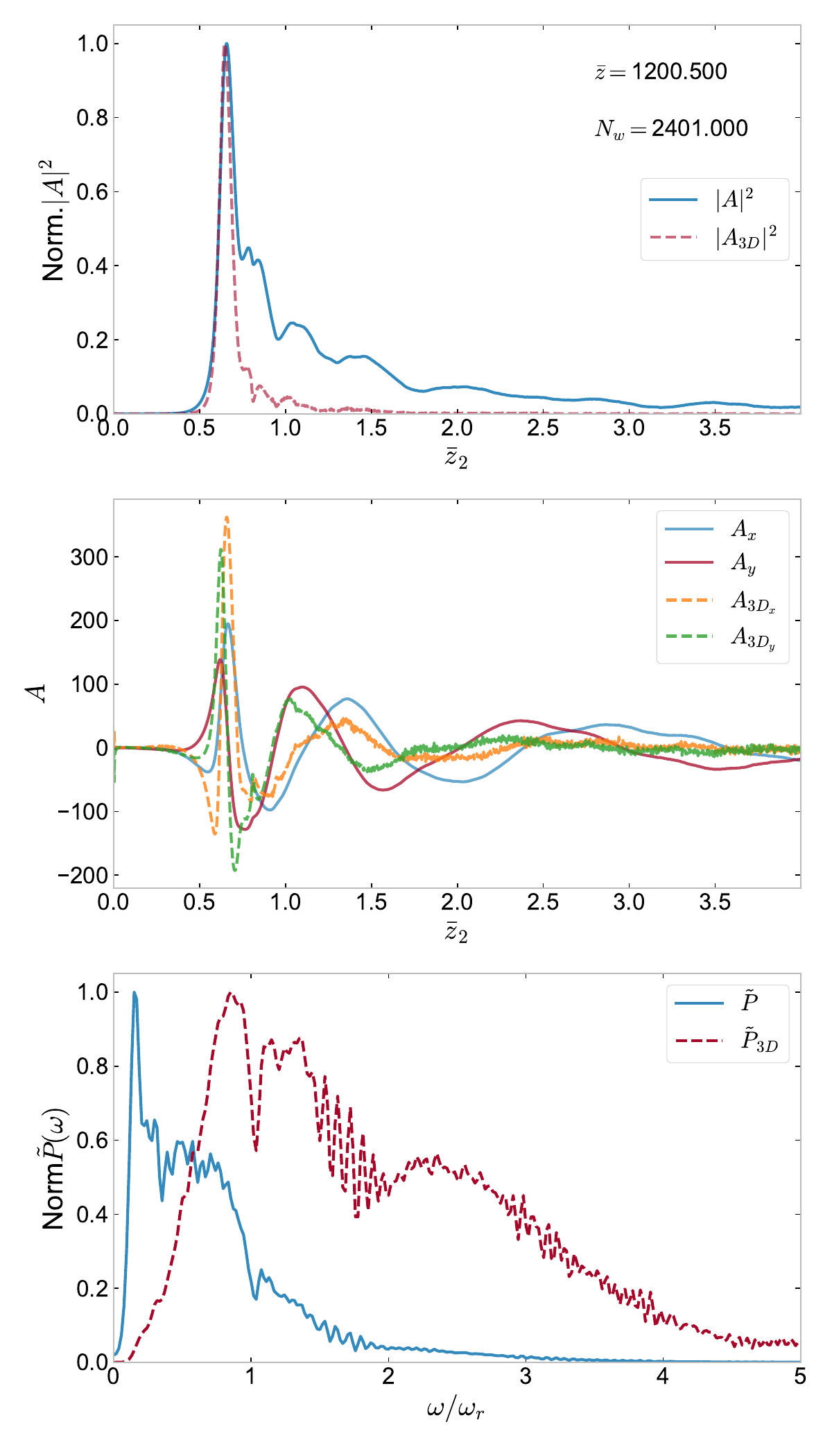}
        \caption{}
        \label{fig:PSD_4pirho=0.5_zbar=1200.5}
    \end{subfigure}
    \begin{subfigure}[c]{0.49\textwidth}
        \includegraphics[width=\textwidth]{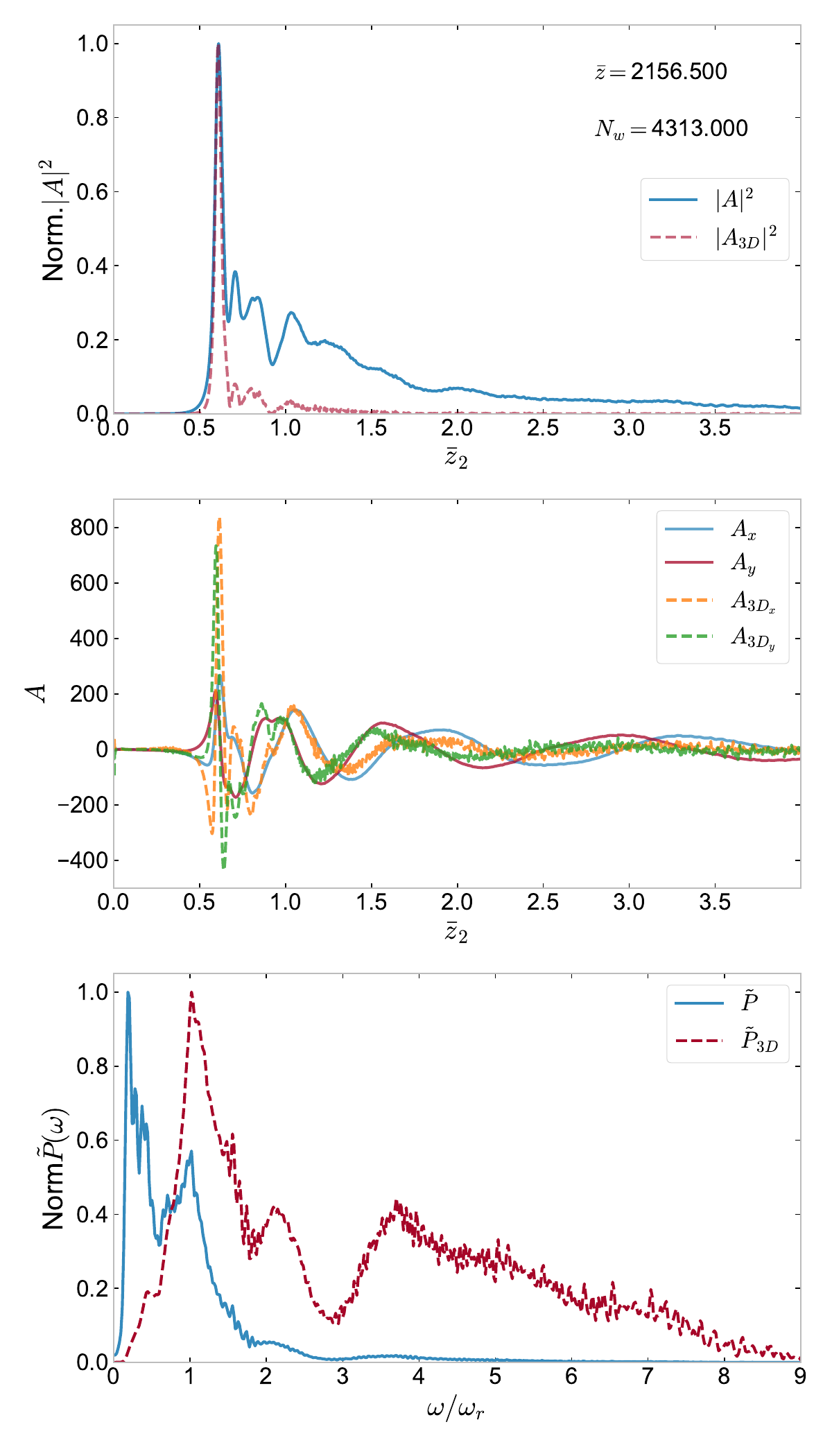}
        \caption{}
        \label{fig:PSD_4pirho=0.5_zbar=2156.5}
    \end{subfigure}
    \caption{As figure~\ref{fig:PSD_4pirho=0.5} but for (a) $\bar{z} = 1200.5$ and (b) $\bar{z} = 2156.5$.}
    \label{fig:PSD_4pirho=0.5_2}
\end{figure*}

These 3D estimates are now shown as the radiation spike evolves through saturation in figure~\ref{fig:PSD_4pirho=0.5} ($\bar{z}=200 \text{ and } 750$) and figure~\ref{fig:PSD_4pirho=0.5_2} ($\bar{z}=1200.5 \text{ and } 2156.5$), for the case where $4\pi\rho = 0.5$. Given the approximations made, these plots cannot be expected to give an accurate numerical estimate, and so are all scaled with respect to their peak values to allow a comparison of their relative temporal and spectral content. 

In figure~\ref{fig:PSD_4pirho=0.5_zbar=200}, for $\bar{z}=200$, the system has entered the superradiant pulse regime and there are similarities between the 3D approximation and the 1D result. The lower frequencies appear to be emitted following propagation of the electrons through the first spike, after $\bar{z}_2\sim 1.5$. This can be inferred also from the PSDs of $\tilde{P}$ and $\tilde{P}_{3D}$ and the electron behaviour of e.g.\ figure~\ref{fig:zb200}, where the electron bunching is spaced at distances larger than one radiation wavelength.  Radiation emitted following the first spike is therefore at a lower frequency, which when reduced due to diffraction, is observed as a reduced $\tilde{P}_{3D}$, compared to the 1D case.
The characteristics of this superradiant pulse emission regime, as derived from the 3D spectrum, show similarities with the experimental results of~\cite{zen2023full}, indicating that the method of introducing 3D diffractive effects above appears not unreasonable.

This behaviour is also demonstrated further in figure~\ref{fig:PSD_4pirho=0.5_zbar=750} for $\bar{z}=750$ which shows that in the 3D approximation, the radiation pulse power following the first spike is reduced, as the lower frequencies emitted by the electrons following the spike diffract away. As the radiation spike  duration decreases, a broader spectrum towards higher frequencies is also observed, and consequently, the radiation field is seen to approach that of a single-cycle pulse.

The above diffractive behaviour is seen to progress well into the saturated regime of the spike evolution as seen in figure~\ref{fig:PSD_4pirho=0.5_2}, for $\bar{z} = 1200.5$ and $2156.5$. 
Despite some small differences in sub-pulse structure, the FWHM of the first peak maintains its sub-resonant wavelength duration sustained by the broad spectral content towards higher frequencies.

\section{Conclusion}
The simulations presented in this paper have provided the first detailed study of how high-power radiation spikes saturate in a FEL. While the simulations are in 1D, and the radiation spikes only interact with a beam of cold electrons, they have revealed at a fundamental level how a superradiant spike saturates, with peak powers orders of magnitude above the normal steady-state value. The saturation process involves the rapid loss of electron energy to the radiation field and their subsequent transit through the spike within one undulator period. The electrons therefore lose their energy to the spike at the sub-wavelength scale. This involves the emission of radiation across a broad range of frequencies well above that of the the fundamental. It was also seen that electrons that enter the spike with an absorptive phase may be rapidly accelerated to much higher energies of $\sim 10$ times or greater than their initial energy.
An examination of the simple scaling of the saturation process agreed well with the numerical results. 

By applying a simplified scaling of radiation diffraction to the 1D radiation field, an estimate of the 3D spectral power of the field was made. This showed, and as observed from the field simulations, that the lower frequency radiation components were emitted after the electrons had passed through the spike and lost energy to it. The spectrum was also similar to that observed in experiment~\cite{zen2023full}.

While the above research will probably not be able to be applied at any FEL facilities in the near future, it does provide further understanding of the fundamental FEL process and may open up new areas of research.

\section{Acknowledgement}
We are grateful to funding from the UKRI Science and Technology Facilities Council (Agreement Number 4163192 Release \#3) and to STFC ST/S006214/1 PWFA-FEL which enabled this research.

\section{Data availability}
Data for the full animation presented in figures~\ref{fig:zb9}),\ref{fig:zb10}),\ref{fig:zb40}),\ref{fig:zb100}),\ref{fig:zb200}),\ref{fig:zb400}),\ref{fig:zb800}) and \ref{fig:zb1100}), is available at \url{https://doi.org/10.15129/60af5807-790e-4a97-9a4b-b0861d8c5fe8}. All other data will be made available on request.

\appendix
\section{Pulse saturation scaling}\label{appendix}
The equations governing the $j$th electron's motion in the scaled radiation frame of reference $\bar{z}_2$ may be written as~\cite{PhysRevA.40.4467}:
\begin{eqnarray*}
\frac{d\bar{z}_{2j}}{d\bar{z}}&=&1-2\rho {{p}_{j}} \\
\frac{d{p}_{j}}{d\bar{z}}&=&-2A(\bar{z}_2)\cos \left( \frac{\bar{z}-{{{\bar{z}}}_{2j}}}{2\rho} \right)
\end{eqnarray*}
where the field $A(\bar{z}_2)$ is assumed constant in $\bar{z}$ and describes a short, high power radiation pulse (spike) of peak power $A_{p}$ into which the electron will propagate.
Assume that the electron starts it interaction with the pulse at resonance, $p_j=0$, and at the phase of maximum rate of energy loss where $\cos \left( \left(\bar{z}-\bar{z}_{2j}\right )/2\rho\right )=1$, then the incremental change in ${p}_{j}$ for a propagation distance $\Delta \bar{z}$ may be written as:
\begin{equation} 
\Delta {p}_j\approx -\frac{A_{p}}{2}\Delta \bar{z},
\end{equation}
where it is assumed $A\approx {A_{p}}/2$ over $\Delta \bar{z}$. 

Similarly, the change in electron position $\Delta \bar{z}_{2j}$ due to its interaction with the radiation field (i.e. not including the resonant drift of $\Delta \bar{z}$ in the radiation frame of reference) may be approximated as:
\begin{equation} 
\Delta \bar{z}_{2j}=-2\rho p_{j}\Delta \bar{z},
\end{equation}
and assuming $p_j\approx \Delta p_{j}/2=-A_{p}\Delta \bar{z}/4$, then:
\begin{equation} 
\Delta \bar{z}_{2j}=\frac{\rho A_{p}}{2}\Delta \bar{z}^{2}
\end{equation} 

Saturation of the electron motion in the radiation pulse is now defined as when the electron propagates an extra half a radiation wavelength through the radiation field, in addition to its drift in the radiation frame of $\bar{z}_2$, in half of a wiggler period. This may be written as  $\Delta \bar{z}_{2j}=\Delta \bar{z}=2\pi \rho$, so that for saturation:
\begin{eqnarray}
2\pi\rho \approx \frac{\rho A_p}{2}(2\pi\rho)^2 \\ 
\Rightarrow  {|A_p|^2} \approx \frac{1}{\pi^2\rho^4}
\label{peakPower}
\end{eqnarray}

The pulse duration of the first peak ${{\tau }_{p}}$, is assumed to scale as the radiation wavelength, which in units of ${{\bar{z}}_{2}}$ is $\tau_p=f\times 4\pi \rho$ where $f$ is a fractional factor. The scaled energy in the first peak at saturation $\varepsilon_p$ is then:
\begin{eqnarray}
\varepsilon_p \approx\tau_p |A_p|^2 \approx  \frac{4f}{\pi\rho^3}
\label{peakEnergy}
\end{eqnarray}





\bibliographystyle{elsarticle-num}
\bibliography{mybib}




\end{document}